\documentclass[a4paper,useAMS,usenatbib]{mnras}

\usepackage{newtxtext,newtxmath}
\usepackage[T1]{fontenc}
\usepackage{ae,aecompl}

\usepackage{graphicx}	
\usepackage{amsmath}	
\usepackage{amssymb}	
\usepackage{setspace}
\usepackage{hyperref}
\hypersetup{draft}
\usepackage{booktabs}   

\newcommand{\FeH} {[\mathrm{Fe}/\mathrm{H}]}

\newcommand{\rrab} {\mbox{RR\emph{ab}}}
\newcommand{\rrc} {\mbox{RR\emph{c}}}
\newcommand{\rrd} {\mbox{RR\emph{d}}}
\newcommand{\typeab} {\mbox{\emph{ab}}}
\newcommand{\typec} {\mbox{\emph{c}}}
\newcommand{\typedd} {\mbox{\emph{d}}}
\newcommand{\Gaia} {\textit{Gaia}}

\newcommand{\Gbp}{G_{\mathrm{BP}}}
\newcommand{\Grp}{G_{\mathrm{RP}}}

\newcommand{\asas}{ASAS-SN-II}

\title[RRL Completeness]{Empirical completeness assessment of the \Gaia~DR2, Pan-STARRS~1 and ASAS-SN-II RR Lyrae catalogues}

\author[C. Mateu et al.]{
Cecilia Mateu,$^{1}$\thanks{E-mail: cmateu@fisica.edu.uy}
Berry Holl,$^{2,3}$
Joris De Ridder,$^{4}$ and
Lorenzo Rimoldini$^{2}$
\\
$^{1}$ Departamento de Astronom\'ia, Facultad de Ciencias, Universidad de la Rep\'ublica, Igu\'a 4225, 14000, Montevideo, Uruguay \\
$^{2}$ Department of Astronomy, University of Geneva, Chemin d'Ecogia 16, 1290 Versoix, Switzerland \\
$^{3}$ Department of Astronomy, University of Geneva, Chemin des Maillettes 51, 1290 Versoix, Switzerland \\
$^{4}$ Institute of Astronomy, KU Leuven, Celestijnenlaan 200D, 3001 Leuven, Belgium
}

\date{Accepted XXX. Received YYY; in original form ZZZ}

\pubyear{2018}

\begin{document}
\label{firstpage}
\pagerange{\pageref{firstpage}--\pageref{lastpage}}
\maketitle

\begin{abstract}
RR Lyrae stars are an important and widely used tracer of the most ancient populations of our Galaxy, mainly due to their standard candle nature. The availability of large scale surveys of variable stars are allowing us to use them to trace the structure of our entire Galaxy, even in previously inaccessible areas like the Galactic disc. In this work we aim to provide an empirical assessment of the completeness of the three largest RRL catalogues available: Gaia DR2, PanSTARRS-1 and ASAS-SN-II. Using a joint probabilistic analysis of the three surveys we compute 2D and 3D completeness maps in each survey's full magnitude range.  
At the bright end ($G<13$) we find \asas~and \Gaia\ are near 100\% complete in \rrab~at high latitude ($|b|>20\degr$); \asas~has the best completeness at low latitude for \rrab\  and at all latitudes for \rrc. 
At the faint end ($G>13$), \Gaia~DR2 is the most complete catalogue for both RRL types, at any latitude, with median completeness rates of $95\%$ (\rrab) and $>85\%$ (RRc) outside the ecliptic plane $(|\beta|>25\degr)$ 
We confirm a high and uniform completeness of PanSTARRS-1 RR Lyrae at 91\% (\typeab) and 82\% (\typec) down to $G\sim18$, and provide the first estimate of its completeness at low galactic latitude ($|b|\leqslant 20\degr$) at an estimated median $65\%$ (\typeab) and $50-60\%$ (\typec). Our results are publicly available as 2D and 3D completeness maps, and as functions to evaluate each survey's completeness versus distance or per line-of sight.
\end{abstract}




\section{Introduction}

The \Gaia~mission in its Second Data Release \citep[DR2][]{GaiaColBrown2018} has provided what is currently the largest, deepest, all-sky catalogue of RR Lyrae (RRL) stars \citep{Clementini2018,Rimoldini2019}, reaching even highly crowded regions in the disc and bulge. The superb astrometry provided by \Gaia~DR2 combined with the power of using RRL stars as standard candles to measure precise distances beyond the reach of \Gaia's parallaxes, are offering an unprecedented view of the structure and kinematics of the old populations ($\geqslant10$~Gyr) of our Galaxy traced by RRLs. \Gaia~DR2 RRLs have allowed studies of the shape of the inner halo \citep{Iorio2018} and it's velocity ellipsoid and gravitational potential in it's innermost regions \citep{Wegg2019}; the identification of new and extra-tidal stars associated to ultra-faint dwarfs \citep{Vivas2020}  and globular clusters \citep{Kundu2019}; new tidal tails around known objects \citet{Minniti2018} and the mapping and kinematic characterisation of known tails like the Sagittarius, Orphan and Pal~5 streams \citep{Ramos2020,Koposov2019,PriceWhelan2019}. 

An aspect shared by these studies is that their results are not sensitive to the selection function and, consequently, do not require a detailed a priori knowledge of the completeness of the catalogue. Other studies do depend critically on this, mainly those that aim at deriving density profiles \citep[e.g.][]{Mateu2018b,Hernitschek2018,Bovy2016}; but also, some techniques aimed to search for stream-like overdensities, which can be prone to yield large rates of false positives if the selection function has sharp variations in completeness \citep[e.g. great circle cell methods,][]{Mateu2011}. 
For the majority of these studies, a detailed map of the completeness as a function of distance for each line of sight is necessary. However, most variable star surveys typically provide completeness estimates based on simulations and can provide only either average estimates, or completeness as a function of distance averaged over the full survey area \citep[e.g.][]{Sesar2017b,Drake2013a,Mateu2012,Vivas2004}. These strategies have proven successful at high latitude, but at low latitudes they are often not-applicable \citep{Sesar2017b} or simplifications need to be made that lead to discarding valuable information at the faint end \citep[e.g.][]{Mateu2018b}. 

A method to empirically estimate the completeness in a statistical manner can overcome these difficulties. Using two independent catalogues --that can be unequivocally cross-matched-- the procedure described by \citet{Rybizki2018}
uses the relative fraction of stars in common between the two to provide the completeness estimates \emph{for both catalogues}; without any prior assumptions on the completeness of either or on their union. To  estimate the completeness of \Gaia~DR2 RRLs we need to combine the other two largest RRL surveys to span
\Gaia's full magnitude range ($G<20.7$) and most of its all-sky coverage:  \asas~covering the full sky at the bright end, down to $G\sim 16$; and Pan-STARRS-1 (PS1), covering 3/4's of the sky at the faint end from $13<G<21$. The joint analysis yields completeness maps for the three catalogues, \Gaia, PS1 and \asas, in their full respective magnitude ranges. These will also be the first measurements of completeness at low latitude ($|b|\leqslant$20\degr) for deep large-scale RRL surveys. Our aim in this work is then to provide an empirical assessment of the completeness of the three largest available RRL catalogues: \Gaia~DR2, Pan-STARRS-1 (PS1) and \asas.

The structure of this paper is as follows. In Section~\ref{s:rrl_catalogues} we describe the three RRL catalogues used: \Gaia~DR2, PS1 and \asas. \Gaia~DR2 published two partially overlapping catalogues containing stars identified as RRLs based solely on their \Gaia~photometry: the Specific Objects Study \citep[SOS,][]{Clementini2018} and VariClassifier \citep[VC,][]{Rimoldini2019} catalogues, produced by two different pipelines and based on data with different characteristics \citep{Holl2018}.  Here we will combine them into a single RRL catalogue VC+SOS. In Section~\ref{s:validation_vcsos} we validate it by comparing it in different aspects to other reference catalogues: we discuss the overall accounting by matching against the PS1 full RRL and bona fide RRL catalogues; the confusion matrix and period recovery statistics comparing against PS1; contamination comparing against the \asas~and Catalina Rapid Transient (CRTS) catalogues of variable stars. In Section~\ref{s:completeness} we compute and discuss the completeness of the three RRL surveys, across the sky and as a function of distance along selected lines of sight. We summarise our conclusions in Section~\ref{s:conclusions}.

\section{The RR Lyrae catalogues}\label{s:rrl_catalogues}

\subsection{\Gaia~VC+SOS}\label{s:GaiaVCSOS}

The second data release of \textit{Gaia} (DR2) included 550\,737 variables stars \citep{Holl2018} amongst which 228\,904 are RRL stars. These RRLs were published in two partially overlapping outputs: (i)~195\,780 sources in the variables classifier \citep[VC, see][]{Rimoldini2019} which contains a label and classification score for each source, and (ii)~140\,784 in the RRL Specific Object Studies \citep[SOS, see][]{Clementini2018}, which contains a large amount of detailed parameters including period, Fourier fitting parameters and photometric metallicity estimates. The overlap between the VC and SOS is 107\,660 stars. Because the SOS output is the result of a more detailed analyses we ignore the VC results for these overlapping sources and only discuss the SOS result. This means that in this paper we examine 88\,120~VC (hereafter `VCnotSOS') and 140\,784 SOS RRLs.

In summary, the notation and \Gaia~RRL samples\footnote{Data publicly 
available in the online \textit{Gaia} archive at \href{http://gea.esac.esa.int/archive/}{\tt http://gea.esac.esa.int/archive/} containing the `tables' and `fields' referred to in the rest of this article} used in this paper are as follows:

\begin{itemize}
    \item VC: \verb+vari_classifier_result+ table with results from the variable classifier \citep{Rimoldini2019}, contains 195\,780 RRLs (with \texttt{best\_class\_name}: \texttt{RRAB}, \texttt{RRC}, \texttt{RRD}, \texttt{ARRD}),
    \item SOS: \verb+vari_rrlyrae+ table from the Specific Objects Study pipeline on RRLs, \citep{Clementini2018} containing 140\,784 SOS RRLs.
    \item VCnotSOS: 88\,120 RRLs in VC table but \emph{not} in SOS.
    \item VC+SOS: 228\,904 RRLs: total of the VC and SOS tables (where SOS results take precedence when there is overlap with the VC).
\end{itemize} 

The first line of Table~\ref{t:qualflag_clean_stats} summarises the initial number of objects in these samples \citep[see also][]{Holl2018}, before the quality cuts described in the following section were applied to filter contaminants out.

\subsubsection{Filtering contaminants with \Gaia~quality flags}

There are contaminant objects both in the VC and SOS RRL tables with much redder colours than normal RRLs. They have \verb+phot_bp_rp_excess_factor+>2 or null, since their $\Gbp$~and $\Grp$ magnitudes are inconsistent with their $G$-band magnitudes. Most, but not all, of these contaminants also have large \verb+astrometric_excess_noise+ (i.e. $\gtrsim2$). Conversely, a small fraction of stars (6,540; i.e. 3.7\%) remain in the clean sample with  \verb+astrometric_excess_noise+$>2$. These stars are kept in the clean sample as their \Gaia~band magnitudes are consistent among themselves and produce distances consistent with their infrared counterparts. They should be culled out, however, in studies aimed at their kinematics.

The 53\,740 stars filtered out this way may be either proper contaminants or they might be real RRLs with contaminated photometry: 12\% of the contaminants are in the LMC and SMC, 80\% are located at low galactic latitude $|b|<20\degr$, out of which $\sim90\%$ are within $50\degr$ of the galactic centre. However, only 67 out of the 53\,740 filtered out contaminants match stars in PS1. This supports that overall these culled out stars are probably not real RRLs since, as we will see in the next section, there is a large overlap between the \Gaia~DR2 VC+SOS and PS1 catalogues.  As noted by \citet{Clementini2018} and \citet{Rimoldini2019}, some of these contaminants (982) turned out to be galaxies mistaken for variable stars. Their \verb+source_id+s are included in Table C.1 of \citet{Clementini2018}.

In what follows we will use the \Gaia~clean sample, even at the expense of a possible loss of completeness, since a loss of completeness does not affect the results at all, but a large contamination would (see Sec.~\ref{s:completeness}).

\begin{table}
\caption{Accounting of the contaminant filtering.}\label{t:qualflag_clean_stats}
\tabcolsep=0.1cm
\begin{footnotesize}
\begin{tabular}{llll}
\toprule
{} &  VC+SOS &           SOS &     VCnotSOS \\
\midrule
Initial      &       228,904 (100\%)  &  140,784 (62\%) &  88,120 (38\%) \\
Contaminants &  53,740 (23\%) 		  &   18,037 (13\%) &  35,703 (41\%) \\
\midrule
Clean        &       175,164 		  &  122,747 (70\%) &  52,417 (30\%) \\
\bottomrule
\end{tabular}
\end{footnotesize}
\end{table}

\begin{figure}
\begin{center}
\includegraphics[width=\columnwidth]{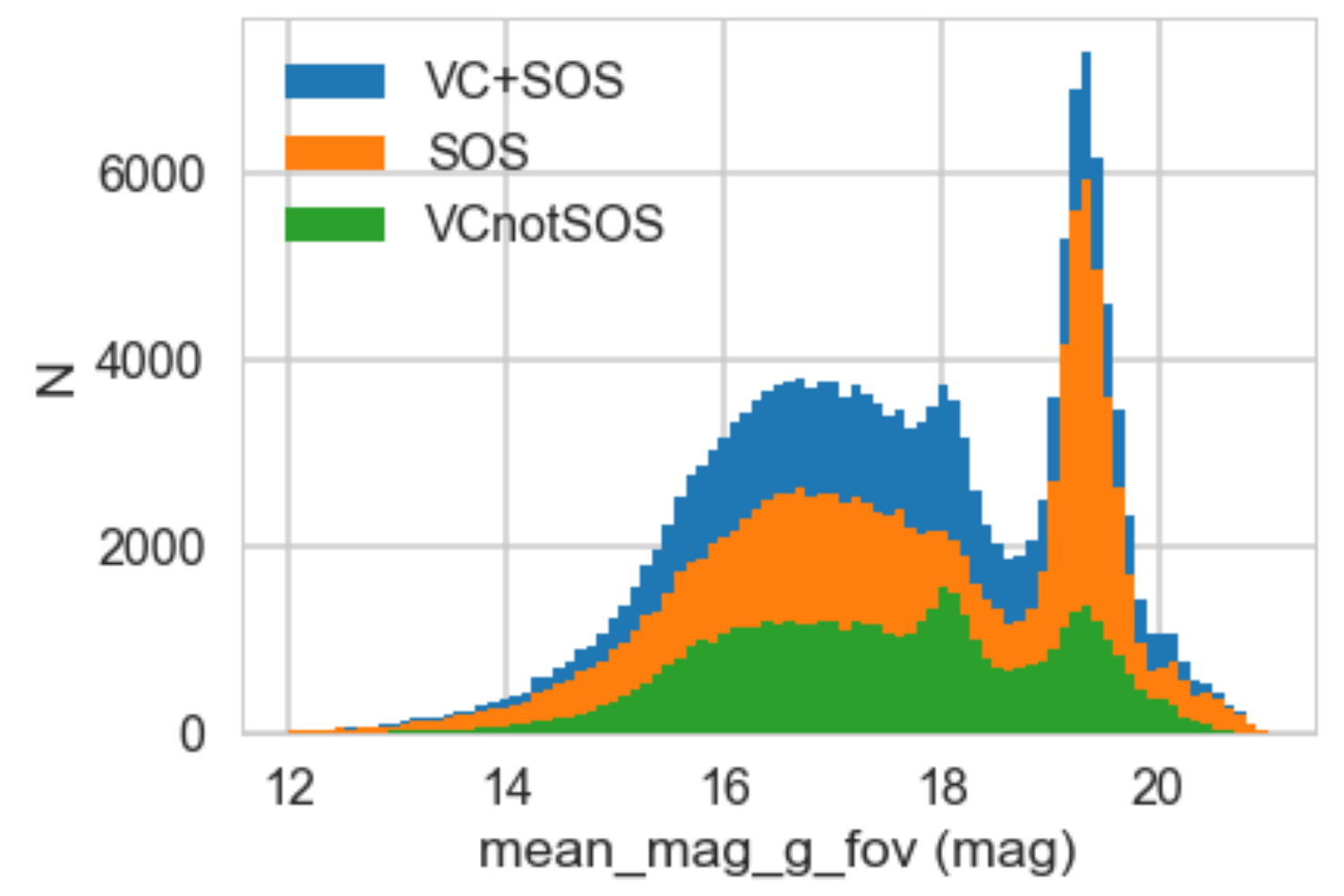}
\caption{Distribution of mean $G$ magnitudes for the (clean) VC+SOS sample and its component SOS and VCnotSOS samples.}\label{f:clean_vcsos_mag_dist}
\end{center}
\end{figure}

The resulting distribution of mean $G$-band magnitudes, \verb+mean_mag_g_fov+ from the \verb+vari_time_series_statistics+ table, is shown in Figure~\ref{f:clean_vcsos_mag_dist} for the clean VC+SOS catalogue we will use hereafter. The peak at $G\sim18$, most prominent in VC+SOS and VCnotSOS but almost not visible in SOS, is mostly due to the main body of the Sagittarius dwarf galaxy. The broad peak at $G\sim19.3$ is due to the Large and Small Magellanic Clouds.

\subsection{Distances}\label{s:distances}

We compute distances to RRLs assuming an absolute magnitude $M_G = 0.6$ for all stars. This corresponds to the absolute magnitude of an RRL with a metallicity $\FeH =-1.5$, the median halo metallicity, according to the $M_G-\FeH$ from \citet{Muraveva2018}. We use this simplified approach since only a subset of stars \citep[$\sim35$\%][]{Clementini2018} are provided in DR2 with a photometric estimate of $\FeH$, and it is the most common way used in the literature to estimate distances for all \Gaia~DR2 RRL stars and we will only use them to provide an estimate of completeness as a function of distance in Sec.~\ref{s:distances}. We do caution the reader, however, these distances will be overestimated for disc RRL, which have higher metallicity (hence, lower luminosity) on average. 

The extinction correction is made taking the $E(B-V)$ from the \citet{Schlegel1998} extinction maps, with the new calibration $A_V=2.742E(B-V)$ from \citet{Schlafly2011}, and converted to $G$-band extinction using the coefficients in eq.~(A.1)  from \citet{Ramos2020} and assuming a dereddened $\Gbp-\Grp=0.7$ for all RRLs (the mean dereddened colour of SOS RRLs).


\subsection{PanSTARRS-1}\label{s:PS1}

The \citet{Sesar2017b} catalogue contains RRL stars identified based on $grizy$ multi-epoch photometry from the Pan-STARRS-1 (PS1) photometric survey spanning $3/4$ of the sky, i.e. the entire sky observable from Hawaii (DEC~$>-30\degr$), with a limiting magnitude ($r\sim21$--$22$) that approximately matches that of \Gaia~$G\sim21$ for the typical colours of RRLs. PS1 $grizy$ light curves are sparsely sampled, typically with $\lesssim 12$ asynchronous epochs per band obtained over a time span of four years. It contains a total of 239,044 \textbf{variable} stars and \citet{Sesar2017b} provide a `classification score' ($\mathrm{score}_{3,ab/c}$) which allows selecting RRLS samples of \typeab~and \typec~with varying degrees of purity and completeness. Selecting \rrab~stars as  $\mathrm{score}_{3,ab}>0.8$ and \rrc~stars    $\mathrm{score}_{3,c}>0.55$, according to \citet{Sesar2017b} results in samples with 0.97 and 0.90 purity respectively, and 0.92 and 0.79 completeness (up to $\sim40$~kpc and at high latitudes) and a total of 44,608 and 17,187 stars for types~\typeab~and \typec~respectively. Therefore, based on their suggested thresholds, for our subsequent analysis we define the two following sub-samples:

\begin{itemize}
    \item PS1 bona fide (61,795 stars): $\mathrm{score}_{3,ab}>0.8$ (\rrab) or   $\mathrm{score}_{3,c}>0.55$ (\rrc)
    \item PS1 non-bona-fide (177,249 stars): the complement of PS1 bona fide
\end{itemize}

We also estimate the $G$-band magnitude for the PS1 RRLs based on their (intensity-averaged) $g$ and $r$ magnitudes, in order to assess what is the overlap and depth of the two catalogues in the G-band and to be able to make broad $G$-band cuts consistently in the two catalogues when assessing the completeness. For this we use the $G-r$ relation given in Table A2 of \citet{Evans2018}. This relation is provided for SDSS filters, we assume they are approximately valid for PS1 filters, for our purposes. Distances are provided for PS1 RRLs by \citet{Sesar2017b}, computed from a $i$-band Period-Luminosity-Metallicity relation (their eq.~5) and assuming $\FeH=-1.5$ for all stars, as we have done for \Gaia~in Sec.~\ref{s:distances}.

\subsection{ASAS-SN-II}\label{s:asas}

\asas~\citep{Jayasinghe2019b,Jayasinghe2019a} is an all-sky catalogue with over half a million classified variable stars\footnote{Light curve parameters and data are available at the \asas~Variable Stars Database \citep{Shapee2014} at \url{https://asas-sn.osu.edu/variables}} spanning the magnitude range $10-11<V\lesssim17$ (corresponding to $G\lesssim 17$ for RRLs, with completeness at $G\sim 16$), including the VSX compilation of variable stars from many different surveys and individual sources. Although the \asas~catalogue includes astrometric data from \Gaia~DR2,  it's variable star identification was done completely independently of \Gaia's, based on it's own optical observations and classification algorithms \citep{Jayasinghe2019a}. 

\begin{figure}
\centering
\includegraphics[width=0.9\columnwidth]{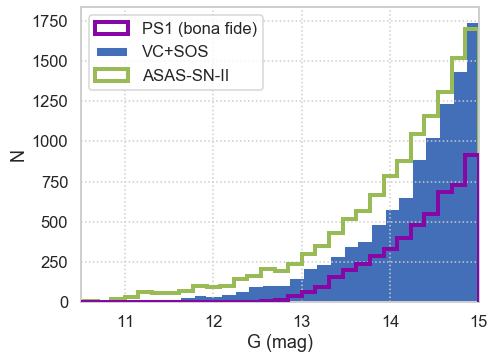}
\caption{G-band magnitude distribution of the ASAS-SN-II RRLs, compared to \Gaia~VC+SOS and PS1 bona fide RRLs.  The bright end tail $G<15$ of the ASAS-SN-II distribution falls off more slowly than \Gaia's at $G<15$, while PS1 falls off more rapidly, going to zero at $G\sim13$.}\label{f:hist_G_asas}
\end{figure}

In total, \asas~\citep{Jayasinghe2019a} contains 44\,279 RRLs of types \typeab, \typec~and \typedd. Here we have restricted the catalogue to the 44\,110 RRLs with periods $<0.95$~d, removing the clear excess of spuriously identified stars with periods of exactly $1$~d (165 stars) and removing also those with suspiciously large amplitudes $>2$~mag (4 stars). The G-band magnitude (obtained for all \asas~stars from \verb+gaiadr2.gaia_source+) distribution of the resulting catalogue of 44\,110 RRLs is shown in Figure~\ref{f:hist_G_asas}, compared to that of \Gaia~and the PS1 bona fide sample.

\subsection{Cross-matches}

For the subsequent analyses we have performed positional cross-matches of \Gaia~DR2 against PS1 and \asas~at tolerances of 3\arcsec and 5\arcsec, respectively. Although at the bright-end ignoring the proper motion could lead to sources being miss-identified, the maximum epoch difference between \asas~and \Gaia~DR2 is small enough (2.5~yr) that this can be ignored when matching at such a large tolerance.

\subsection{Contamination}\label{s:contamination}

As we will see in Sec.~\ref{s:completeness}, the method we will use to estimate completeness relies on the assumption that the catalogue's contamination rates are small enough to be negligible. Here we assess the three RRL catalogue's contamination rates. 

For the \Gaia~VC+SOS RRL catalogue we can estimate the contamination by other types of variable stars using the \asas~survey, which in addition to RRLs has identified many types of period variables. Cross-matching \asas~to VC+SOS with a 5" tolerance\footnote{Following  \citep{Jayasinghe2019a}.} yields  32\,092 matching stars out of which 1\,902 are classified as variables other than RRLs.
This number represents $\sim6\%$ of the RRLs present in VC+SOS in the \asas~footprint (all-sky) up to its completeness magnitude $G\sim16$. When calculated separately for each RRL type, the contaminant fraction is slightly smaller for \rrab~($5\%$) and higher for \rrc~($9\%$), as is usually the case since the latter are more prone to confusion with other types of variable star (e.g. contact eclipsing binaries or $\delta$ Scuti stars). A similar mean contaminant fraction of $7\%$ is obtained when the same procedure is applied using the Catalina Rapid Transient Survey catalogues of Periodic Variable stars from \citet{Drake2014,Drake2017}. These two results are very similar and only slightly higher than those of \citet{Holl2018}, who for this magnitude range estimate it at $\sim5\%$. Their estimate is based on a random sample of SOS RRLs in the full magnitude range ($G<20.7$), yielding a $9\%$ contamination in total, of which $\sim4\%$ is due to the faintest stars $20<G<20.7$. 

For the PS1 catalogue, \citet{Sesar2017b} estimated the purity in 91\%  and 90\% respectively for bona-fide \rrab\ and \rrc\ stars, down to the survey's limiting magnitude. In both cases, then, contamination is expected to be $\sim9\%$. They obtained this estimate by comparing against the SDSS Stripe 82 catalogues found to be 100\% pure and complete for RRLs \citep{Sesar2010,Suveges2012}. When we repeat the above procedure by comparing against the \asas~survey, we find contamination rates of $2\%$ and $15\%$ for \rrab\ and \rrc\ stars, at $|b|>20\degr$, the latitude range of the Stripe 82 catalogue.  This estimate, although slightly lower for \rrab\ and higher for \rrc, is in better agreement with our expectations that contamination rates for \rrc~stars are usually higher than those for \rrab. The discrepancy between the two estimates probably stems from the different procedures: ours is based on the cross-identification with a catalogue in which periodic variable stars have been  classified based on several attributes, while the assertion that the Stripe 82 catalogue is free of contamination is based solely on the visual inspection of the RRL light curves and their quality. 
That, combined with the more extensive area coverage of the \asas~catalogue, leads us to expect our estimate to be more representative of the contamination of the PS1 survey as a whole. Finally, we find the contamination of the full PS1 survey to be $10\%$ on average, and respectively $3\%$ and $23\%$ for \rrab~and \rrc stars. As expected, the contamination is higher in the areas closer to the Galactic disc, particularly for \rrc~stars.  

Finally, for the \asas~catalogue, we use the same procedure as for \Gaia~and compare against the Catalina Rapid Transient Survey catalogues of Periodic Variable stars from \citet{Drake2014,Drake2017}, matching again at a tolerance of 5". In the area of overlap between the two surveys ($\mathrm{DEC}<70\degr$ and $|b|>30\degr$) we find a $9\%$ contamination in total for RRLs of both types, down to $G<16$, corresponding to very similar rates of $9\%$ and $8\%$ respectively for RRLs of type \typeab~and \typec.  

\section{Validation of the VC+SOS catalogue}\label{s:validation_vcsos}

In what follows, most of the validation is done against PS1 because of its large area coverage ($3\pi$ of the sky) and depth comparable to \Gaia. However, in Sec.~\ref{s:contamination} we compare against the catalogues of periodic variable stars of the \asas~survey to estimate the fraction of contaminants present in the VC+SOS catalogue. 

\subsection{Overall validation against PS1}\label{s:ps1_validation}

\subsubsection{Accounting}\label{s:accounting}
\begin{figure}
\begin{center}
\includegraphics[width=\columnwidth]{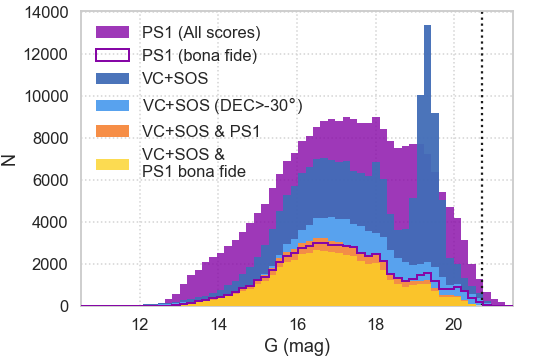}
\caption{$G$-band (intensity-averaged) magnitude distribution of different matching sub-samples of the PS1 and VC+SOS RRL catalogues}\label{f:vcsos_ps1_gmag_hist}
\end{center}
\end{figure}

\begin{figure*}
\begin{center}
\includegraphics[width=2\columnwidth]{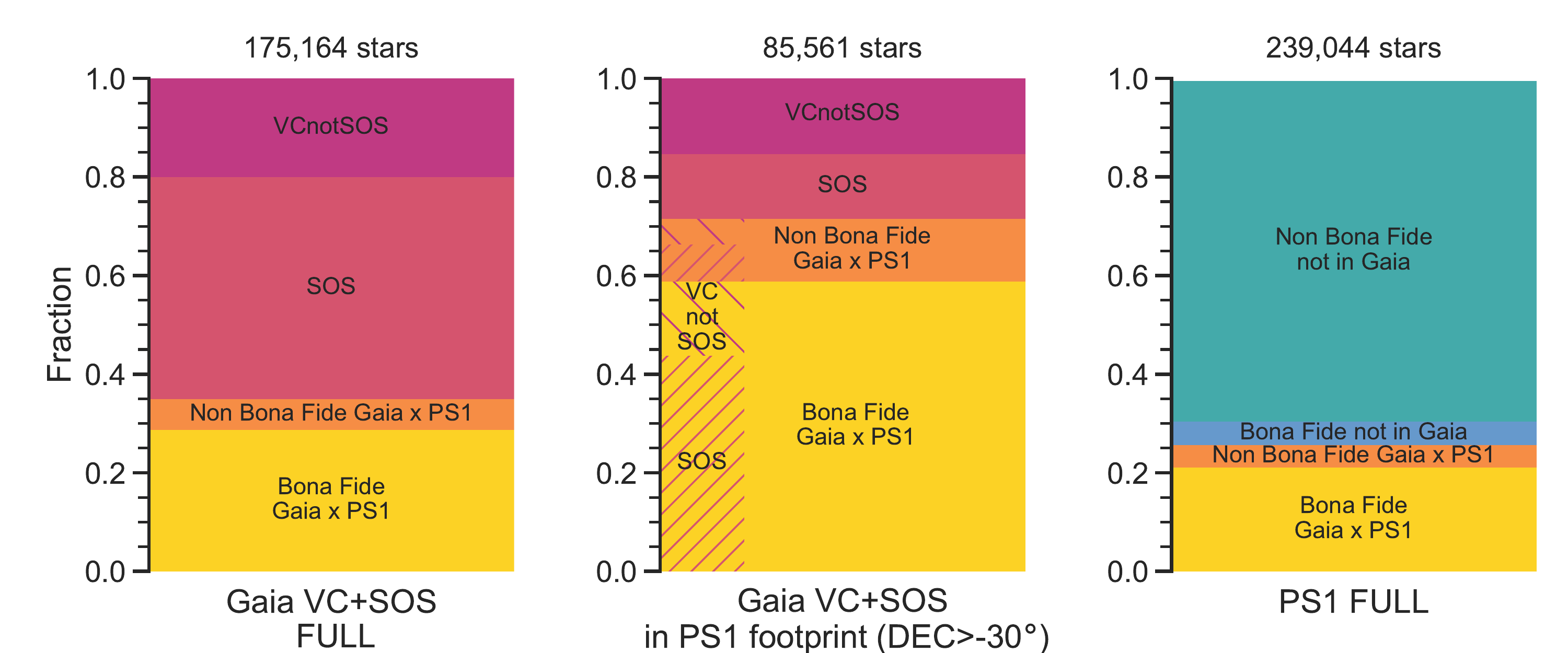}
\caption{Accounting: fraction of \Gaia~and PS1 matching RRLs (Gaia $\times$ PS1), separated as SOS and VCnotSOS for \Gaia, and as bona fide and non-bona fide for PS1. \emph{Left:} for the full \Gaia~VC+SOS sample. \emph{Center:} for the \Gaia~VC+SOS stars in the 3$\pi$ PS1 footprint (DEC~$>30\degr$). \emph{Right:} for the full PS1 sample. }\label{f:accounting}
\end{center}
\end{figure*}

Figure~\ref{f:vcsos_ps1_gmag_hist} shows the $G$-band magnitude distribution for the PS1, VC+SOS and different matching sub-samples. In order to show the PS1 RRL data in this plot, we compute the expected $G$-band magnitude for each PS1 star based on its  $g$ and $r$ intensity-averaged magnitudes (observed, i.e. without extinction correction), using the $G(g,g-r)$ relation from \citet[][Appendix~A]{Evans2018}. Figure~\ref{f:accounting} summarises the composition in the different sub-samples (SOS, VCnotSOS, PS1 bona fide, PS1 non-bona-fide) the full \Gaia~VC+SOS catalogue (\emph{left}); the \Gaia~VC+SOS sub-catalogue overlapping the PS1 footprint (\emph{middle}), i.e. restricted to DEC~$>-30\degr$; and the full PS1 RRL catalogue (\emph{right}). 


Figure~\ref{f:vcsos_ps1_gmag_hist} shows that, for RRLs, PS1 has a similar limiting magnitude as \Gaia.  Out of the 239K total variables in PS1, 61\,208 (~26\%) match with \Gaia~VC+SOS objects, even though all of them are potential matches given that the two surveys have a similar limiting magnitude. Although the majority of the 239K total stars are not expected to be true RRLs, we are interested in analysing these cross-match statistics to check how many true RRLs may be left behind in the non-bona fide sample. Out of the 61K matches between \Gaia~VC+SOS and PS1, 82\% (50\,298) are bona fide PS1 RRL. This, together with the previous point, suggests the vast majority of stars in the 177K non-bona fide PS1 catalogue are indeed not RRL stars. Conversely, there are 11K non-bona fide stars with an RRL counterpart in \Gaia; as we will see in our analysis of Period recovery in Sec.~\ref{s:period_recovery}, these are probably genuine RRL that simply did not make the bona fide cut in PS1. Hence, it would be useful to analyse these star's PS1 time series in combination with \Gaia's.  Finally, the 50K matching RRL constitute 81\% of the total PS1 bona fide sample. So, most of the (VC+SOS)xPS1 matches are bona fide (82\%) RRL in PS1 and most (81\% also) of the bona fide PS1 RRL are recovered.

The (VC+SOS)xPS1 matching RRL amount to 70\% of the objects in the VC+SOS sample at DEC~$>-30\degr$, i.e. in the area overlapping the PS1 footprint where a match is possible. Since roughly two-thirds of the matching objects are bona fide PS1 RRL, this means just over half (57\%) the objects in VC+SOS (inside the PS1 footprint) are bona fide PS1 RRL. The question is, in the VC+SOS (DEC~$>-30\degr$) catalogue, how confident are we in the remaining 43\%. These can be classified in two classes:  the ones that  are in PS1, but are not bona fide; and the ones not in PS1. To check this, we will analyse the period recovery stats for both the bona fide and non-bona-fide samples in Section~\ref{s:period_recovery}. The left panel of Fig.~\ref{f:accounting} also clearly shows that PS1 RRL only represent $<40\%$ of the \Gaia~RRL, even though PS1 covers 3/4 of the sky at the same limiting magnitude as \Gaia~and the recall of PS1 RRL by \Gaia~is large, so roughly $>60\%$ of the RRLs in \Gaia~VC+SOS are new discoveries. The main reasons for this are: that the LMC and SMC contain thousands of RRLs and are not in PS1's footprint; and that PS1's completeness drops of very rapidly at low Galactic latitude due to crowding, where \Gaia~has a much better performance, as we will see in Section~\ref{s:completeness_ps1}.
    
\subsubsection{Period Recovery}\label{s:period_recovery}

\begin{figure}
\begin{center}
\includegraphics[width=0.5\columnwidth]{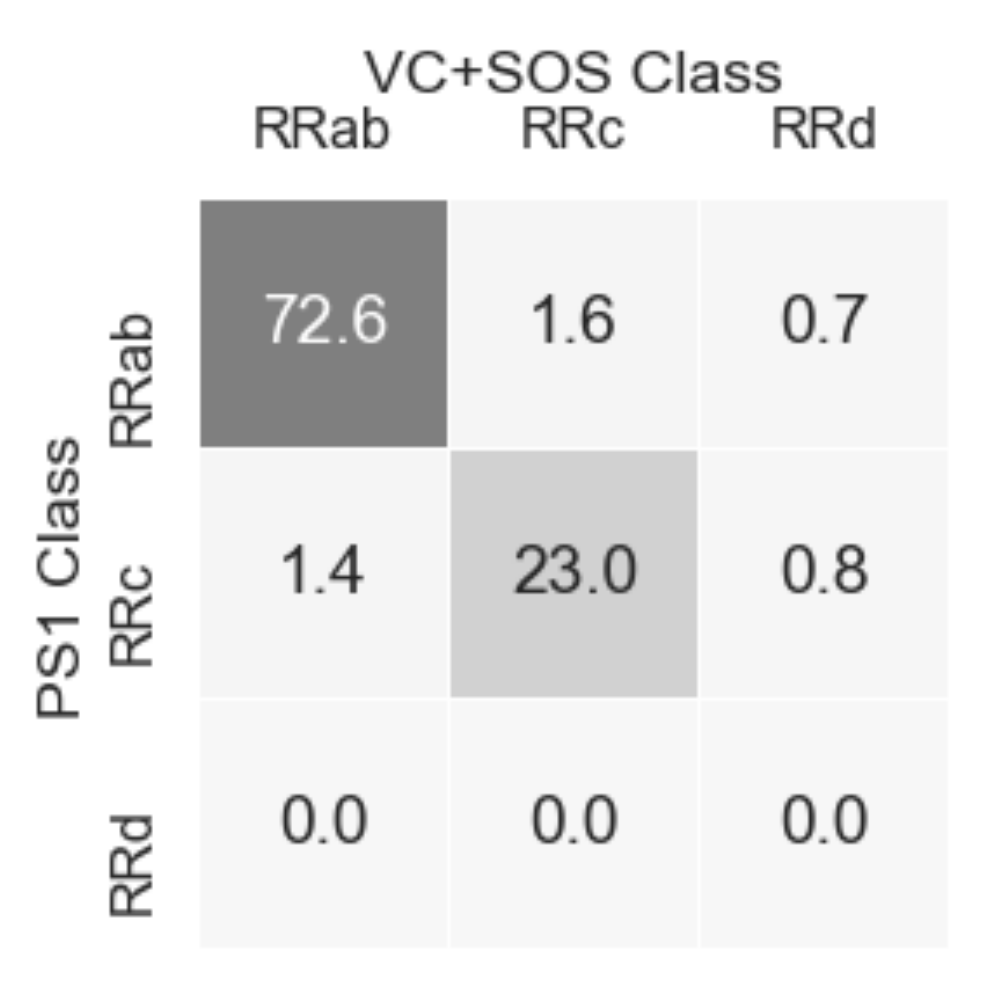}
\caption{Confusion matrix for (VC+SOS)xPS1}\label{f:vcsos_ps1_confusion}
\end{center}
\end{figure}

First, we cross-check against PS1 the classification of VC+SOS RRL into the three types, by computing the confusion matrix, shown in Figure~\ref{f:vcsos_ps1_confusion}. This shows the overall class recovery is good, $>95\%$ of the RRL stars are recovered with the same classification in both surveys. We note that this is particularly good since we are not discriminating by either VCnotSOS/SOS or bona-fide/non-bona-fide and gives us confidence that both the SOS and VC classes are indeed reliable. 

\begin{figure}
\begin{center}
\includegraphics[width=0.99\columnwidth]{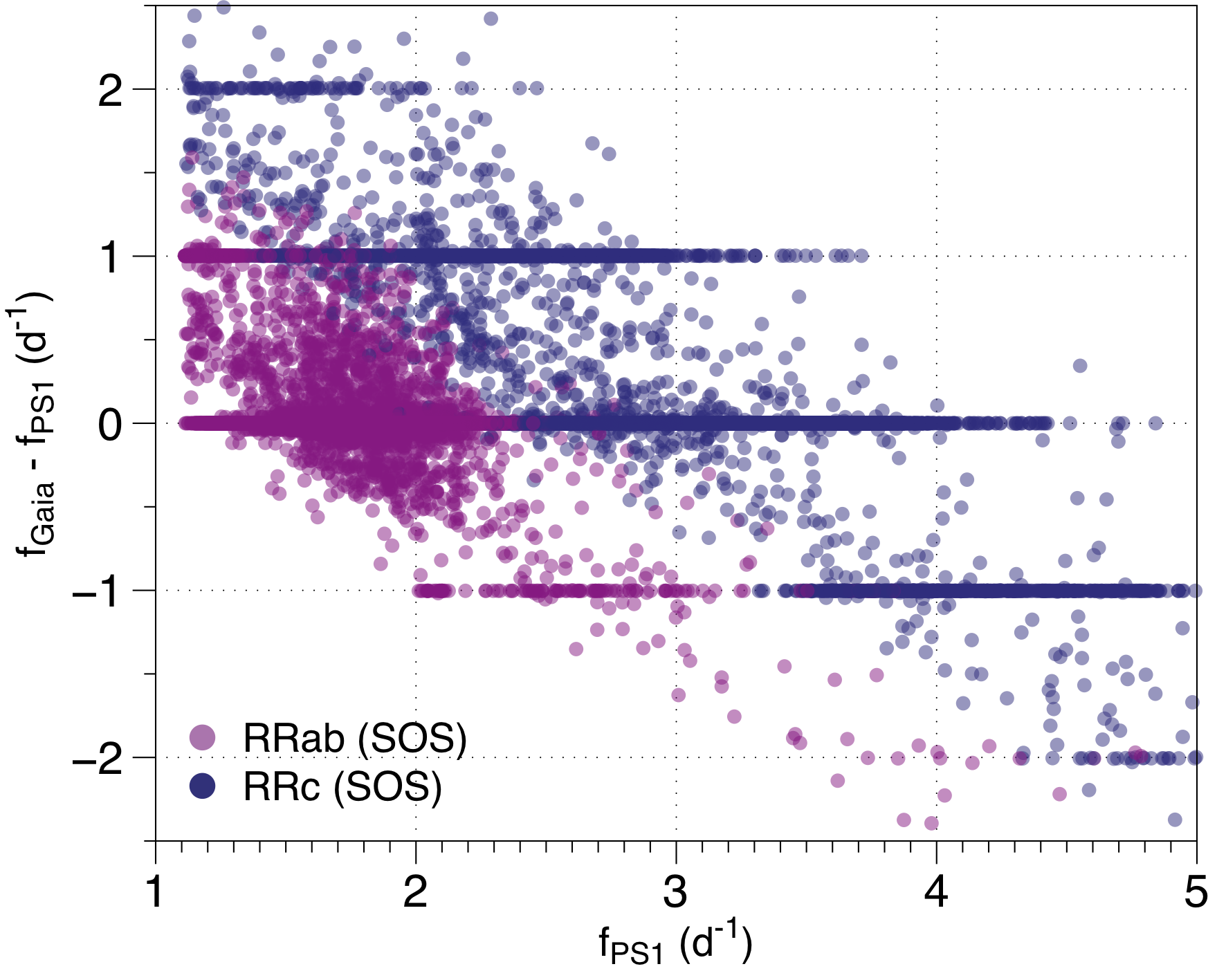}
\includegraphics[width=0.99\columnwidth]{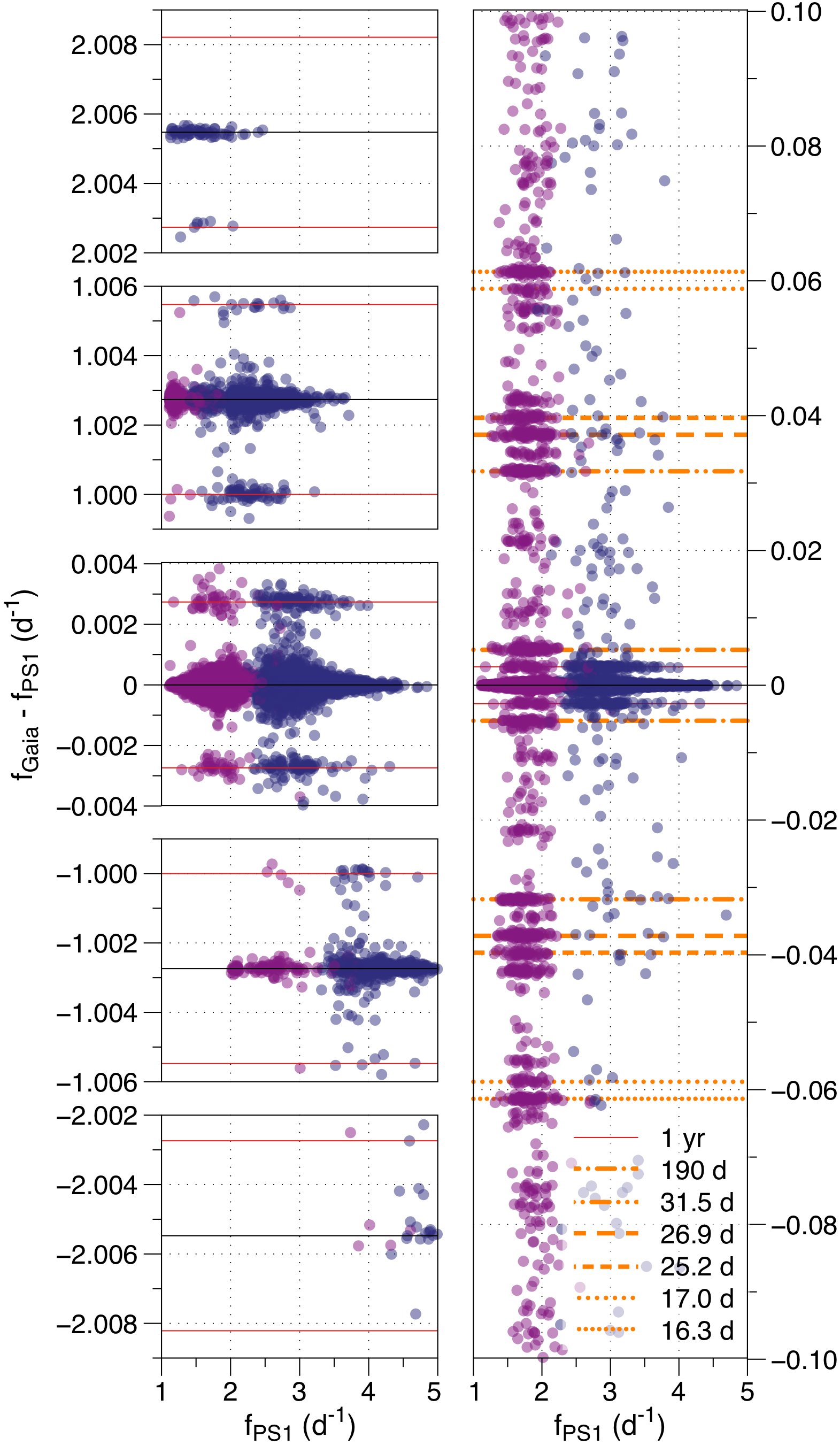}
\caption{\Gaia~DR2 versus PS1 periods for matching RRLs. The identity and main period alias loci are shown with the solid and different dashed lines, respectively. RRL classified as type \typeab~and \typec~(SOS) are shown with different colours. The lower left panel is a zoomed-in version of the 0, $\pm1$ and $\pm2~\text{d}^{-1}$, and the lower right an expansion around 0, showing a richness of various aliases (mainly from \Gaia). The main period aliases were visually identified as the over-dense loci in this plot.
}\label{f:period_rec}
\end{center}
\end{figure}

The period recovery comparison between PS1 and VC+SOS can only be performed for the SOS subset, since the VCnotSOS RRL do not have reported periods. The top panel of Figure~\ref{f:period_rec} shows the \Gaia~DR2 versus PS1 frequency, with the bottom panel zooming in on the 1 and 0.5~day aliases, as well as the \Gaia~aliases around the central part around $f_\mathrm{Gaia} - f_\mathrm{PS1}=0$. In addition to the identity line, the over-dense horizontal lines where many stars align in this plot correspond to aliased periods, which typically show up when one, or more, external periodicities are embedded in the time sampling.  Table~\ref{t:period_rec} summarises the fraction of stars recovered with the same period or an alias, within a tolerance of $X=3$ and $10$ `sigma' on the normalised frequency difference  (see Eq.~\ref{eq:aliasCorrectedRecovery} of Appendix~\ref{a:perRecovCrit}). These two criteria can be considered as `strict' and `loose' period recovery criteria, respectively. For example, for an RRab star with true period $P=0.6$~d, the 3 and 10 sigma criteria 
correspond to a 6.0 and 19.8~sec period tolerance, and for an RRc star with true period $P=0.3$~d, the 3 and 10 sigma criteria correspond to a 1.8 and 5.9~sec period tolerance (using Eq.~\ref{f:deltaPeriodEquivalent}). See Figs.~\ref{f:diffFhist} and \ref{f:normFreqDiffDistr} for a visual representation of these tolerances with respect to the data. 

Table~\ref{t:period_rec} shows that the vast majority of the RRLs are recovered with the same period: for the full SOSxPS1 sample, more than 77\% of the stars are recovered within 3-sigma, and about 83\% within 10-sigma. For RRLs it is well known that period aliasing is a relatively common issue since certain common external periodicities, e.g. 1~d, produce period aliases that are also in the typical range of RRL periods \citep{Lafler1965}. 
As Figure~\ref{f:period_rec} and Table~\ref{t:period_rec} show, the 1~d alias and its harmonic the 0.5~d alias are the most common ones and affect mostly \rrc~stars (blue dots), as is typically the case \citep[see e.g.][]{Mateu2012,Vivas2004}.
These are associated with PS1's ground-based observation cadence.
Other external periodicities are revealed too, e.g.: 16,3~d, 17~d, 25~d, 27~d, 31~d, 190~d, 1~yr and various others. These less common ones are a result of \Gaia's observation cadence induced by its specific scanning law.

In Table~\ref{t:period_rec}, we show also the statistics for the SOS matching RRLs classified as bona fide and non-bona-fide in PS1 (see Section~\ref{s:ps1_validation}). We do this as a way to get an estimate of how likely it is that the non-bona-fide stars in common between \Gaia~VC+SOS and PS1 are actual RRLs, given that they are a small but non-negligible fraction (~10\%) of the matching stars. The table shows that about 46\% of the non-bona fide stars are recovered at 3 sigma, increasing to over half (54\%) when adopting the looser 10 sigma tolerance.
The fraction of stars recovered at the 1~d alias relative to the identity is also much higher, having increased by a factor of $\sim4$ compared to the bona~fide sample. 
All of this suggests the non-bona-fide stars with SOS counterparts could be legitimate RRL stars with either noisier or more scarcely sampled light curves for which a period recovery is more prone to aliasing.

\begin{table}
\caption{Percentage of RRLs (\typeab~and \typec) with periods recovered correctly or as period aliases, using the `sigma' recovery criterion of Eq.~(\ref{eq:aliasCorrectedRecovery}). }\label{t:period_rec}
\begingroup
\setlength{\tabcolsep}{4.3pt} 
\begin{tabular}{lrrrrrr}
\toprule
{}& \multicolumn{2}{c}{All} & \multicolumn{2}{c}{bona fide}  & \multicolumn{2}{c}{non-bona-fide} \\
{$X$ `sigma' recovery} & ($3$)  & ($10$) & ($3$)  & ($10$) & ($3$)  & ($10$)  \\
\midrule

\multicolumn{7}{l}{RRab + RRc} \\
Identity                   &   77.35    &   83.08    &    82.67    &    88.05   &   45.89    &    53.66   \\
$\pm$2.0055 d$^{-1}$ (0.5~d)  &   0.17    &   0.19    &    0.03    &    0.03   &   1.03    &    1.11   \\
$\pm$1.0027 d$^{-1}$ (1.0~d)  &   6.56    &   7.20    &    4.74    &    5.05   &   17.33    &    19.91   \\
$\pm$0.0613 d$^{-1}$ (16.3~d)  &   0.13    &   0.28    &    0.15    &    0.31   &   0.06    &    0.08   \\
$\pm$0.0590 d$^{-1}$ (16.9~d)  &   0.03    &   0.09    &    0.04    &    0.11   &   0.00    &    0.02   \\
$\pm$0.0397 d$^{-1}$ (25.2~d)  &   0.09    &   0.23    &    0.09    &    0.25   &   0.03    &    0.10   \\
$\pm$0.0372 d$^{-1}$ (26.9~d)  &   0.19    &   0.40    &    0.19    &    0.40   &   0.22    &    0.42   \\
$\pm$0.0317 d$^{-1}$ (31.5~d)  &   0.26    &   0.37    &    0.27    &    0.39   &   0.16    &    0.24   \\
$\pm$0.0053 d$^{-1}$ (190.0~d)  &   0.21    &   0.41    &    0.21    &    0.40   &   0.19    &    0.43   \\
$\pm$0.0027 d$^{-1}$ (1~yr)  &   0.48    &   0.65    &    0.36    &    0.46   &   1.17    &    1.80   \\
\midrule
Total  &   85.47   &   92.89   &   88.75   &   95.45   &   66.09   &   77.76  \\
\bottomrule

\multicolumn{7}{l}{RRab} \\
Identity                   &   83.78    &   89.71    &    86.67    &    92.13   &   61.92    &    71.39   \\
$\pm$2.0055 d$^{-1}$ (0.5~d)  &   0.00    &   0.01    &    0.00    &    0.00   &   0.03    &    0.11   \\
$\pm$1.0027 d$^{-1}$ (1.0~d)  &   1.25    &   1.49    &    0.43    &    0.49   &   7.49    &    9.02   \\
$\pm$0.0613 d$^{-1}$ (16.3~d)  &   0.18    &   0.38    &    0.20    &    0.41   &   0.08    &    0.11   \\
$\pm$0.0590 d$^{-1}$ (16.9~d)  &   0.05    &   0.13    &    0.05    &    0.14   &   0.00    &    0.03   \\
$\pm$0.0397 d$^{-1}$ (25.2~d)  &   0.12    &   0.31    &    0.13    &    0.33   &   0.06    &    0.14   \\
$\pm$0.0372 d$^{-1}$ (26.9~d)  &   0.25    &   0.53    &    0.24    &    0.52   &   0.34    &    0.65   \\
$\pm$0.0317 d$^{-1}$ (31.5~d)  &   0.35    &   0.51    &    0.36    &    0.52   &   0.28    &    0.37   \\
$\pm$0.0053 d$^{-1}$ (190.0~d)  &   0.29    &   0.53    &    0.29    &    0.53   &   0.28    &    0.51   \\
$\pm$0.0027 d$^{-1}$ (1~yr)  &   0.18    &   0.28    &    0.12    &    0.16   &   0.68    &    1.13   \\
\midrule
Total  &   86.46   &   93.87   &   88.49   &   95.25   &   71.16   &   83.46  \\
\bottomrule

\multicolumn{7}{l}{RRc} \\
Identity                   &   62.18    &   67.43    &    72.11    &    77.28   &   24.89    &    30.44   \\
$\pm$2.0055 d$^{-1}$ (0.5~d)  &   0.57    &   0.60    &    0.10    &    0.12   &   2.33    &    2.41   \\
$\pm$1.0027 d$^{-1}$ (1.0~d)  &   19.08    &   20.68    &    16.11    &    17.08   &   30.22    &    34.19   \\
$\pm$0.0613 d$^{-1}$ (16.3~d)  &   0.02    &   0.04    &    0.01    &    0.04   &   0.04    &    0.04   \\
$\pm$0.0590 d$^{-1}$ (16.9~d)  &   0.00    &   0.02    &    0.00    &    0.02   &   0.00    &    0.00   \\
$\pm$0.0397 d$^{-1}$ (25.2~d)  &   0.01    &   0.04    &    0.01    &    0.04   &   0.00    &    0.04   \\
$\pm$0.0372 d$^{-1}$ (26.9~d)  &   0.05    &   0.09    &    0.05    &    0.09   &   0.07    &    0.11   \\
$\pm$0.0317 d$^{-1}$ (31.5~d)  &   0.04    &   0.05    &    0.05    &    0.05   &   0.00    &    0.07   \\
$\pm$0.0053 d$^{-1}$ (190.0~d)  &   0.02    &   0.11    &    0.00    &    0.05   &   0.07    &    0.33   \\
$\pm$0.0027 d$^{-1}$ (1~yr)  &   1.17    &   1.53    &    1.00    &    1.22   &   1.81    &    2.67   \\
\midrule
Total  &   83.13   &   90.58   &   89.44   &   95.99   &   59.44   &   70.30  \\
\bottomrule
\end{tabular}
\endgroup
\end{table}

\section{Completeness}\label{s:completeness}

We estimate the completeness of two independent catalogues following the procedure used by \citet{Rybizki2018}. In our case we will take VC+SOS and \asas~or VC+SOS and PS1 as the two independent RRL catalogues.  

The procedure is as follows. Assuming there is no (or negligible) contamination, the completeness $C_A$ of survey $A$ is given by its probability of detection of a star which is, in turn, the number $N_A$ of stars observed by the survey divided by the true, total, number of stars $N_{\mathrm{true}}$:

\begin{eqnarray}
    C_A \equiv P_{A} = N_{A}/N_{\mathrm{true}} \label{e:Cgdef} 
\end{eqnarray}

Assuming that two surveys are conditionally independent ($P_{A\cap B}=P_A\cdot P_B$), the number $N_{A\cap B}$ of stars in common between the two is given by:

\begin{equation}
N_{A\cap B} = P_{A\cap B}N_{\mathrm{true}} = C_A C_B N_{\mathrm{true}} \label{e:Ngp}
\end{equation}
which, substituting Eq.~(\ref{e:Cgdef}) for survey A and similarly for survey B, and solving for the completeness, gives the two following expressions: 

\begin{eqnarray}
    C_A = N_{A\cap B}/N_B \label{e:Cg}\\
    C_B = N_{A\cap B}/N_A \label{e:Cp}
\end{eqnarray}

The validity of this procedure to estimate the two survey's completeness, therefore, hinges on two assumptions: that the surveys are conditionally independent, as already mentioned, and that there is no contamination. In Sec.~\ref{s:contamination} we estimated contamination to be $<10\%$ for the three surveys, we will assume in what follows that the approximation of negligible contamination is valid. The method also makes no assumption on the completeness of either individual survey or on the combination ($A \cup B$) of the two.

The method also implicitly assumes the cross-match between the two catalogues unequivocally identifies an object present in one survey with its correct counterpart in the other. Although this can be challenging in crowded fields when matching two surveys with very different spatial resolution, like \Gaia\  and \asas, it is not a problem with RRLs. In the \Gaia~VC+SOS catalogue only 42(19) RRLs have another star at $<$5\arcsec($<$3\arcsec), which could result in a dubious cross-match against \asas(PS1). We can safely ignore these as they will have no impact in our calculations.

\subsection{Completeness at the bright end: \Gaia~VC+SOS and ASAS}\label{s:completeness_asas}

\begin{figure*}
\begin{center}
\includegraphics[width=2\columnwidth]{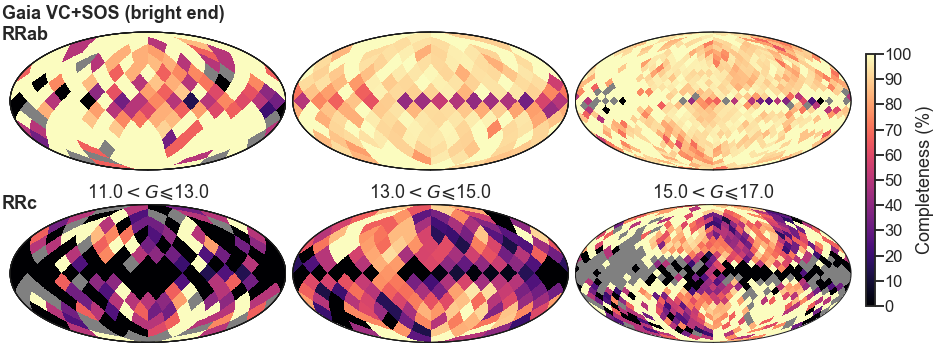}
\caption{Completeness maps for VC+SOS \rrab~(\emph{top}) and \rrc~or \rrd~(\emph{bottom}) stars, estimated from matches agains \asas, in three G-band magnitude ranges. All maps are Mollweide projections in Galactic coordinates, with $l=0\degr$ at the centre and longitude increasing to the left. The gray shading denotes pixels with undefined completeness (i.e. zero counts in the comparison survey).}\label{f:gaia_completeness_G_bright_maps}
\end{center}
\end{figure*}

\begin{figure*}
\begin{center}
\includegraphics[width=2\columnwidth]{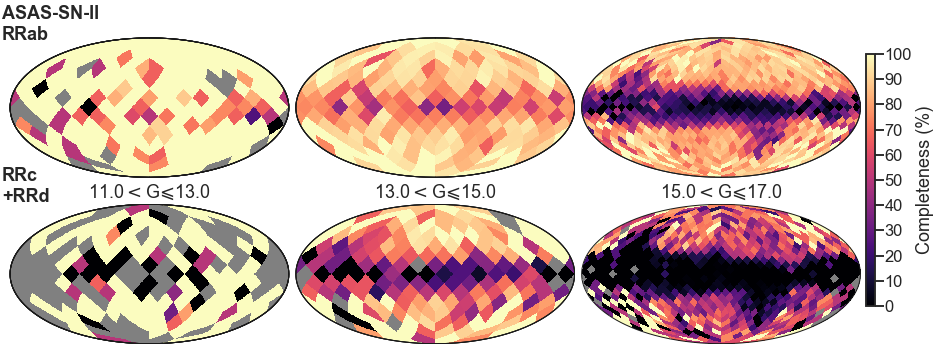}
\caption{Completeness maps for \asas~ \rrab~(\emph{top}) and \rrc~or \rrd~(\emph{bottom}) stars, estimated from matches against \Gaia~VC+SOS, in three G-band magnitude ranges. All maps are Mollweide projections in Galactic coordinates, with $l=0\degr$ at the centre and longitude increasing to the left. The gray shading denotes pixels with undefined completeness (i.e. zero counts in the comparison survey).}\label{f:asas_completeness_G_bright_maps}
\end{center}
\end{figure*}

First we  take the \Gaia~VC+SOS and \asas~RRL catalogues as surveys A and B in Eqs.~(\ref{e:Cg}) and (\ref{e:Cp}) to estimate each catalogue's completeness. There estimates are valid in the magnitude range $11\leqslant G\leqslant17$, where the two surveys overlap.  Figures~\ref{f:gaia_completeness_G_bright_maps} and~\ref{f:asas_completeness_G_bright_maps}  show the resulting completeness maps for \Gaia~VC+SOS and \asas, respectively, separately for the two RRL types: \typeab~(top row) and \typec~(bottom row). 

For \Gaia~VC+SOS the completeness is remarkably high ($>80\%$) and homogeneous across the sky for the \typeab, even close to the Galactic plane. For the \typec~completeness is systematically lower on average than for the \typeab. Although at the faintest end (right panels) both types start to show the \Gaia~scanning law pattern, its clear the \rrc\  are much more affected by it. These results are to be expected: \rrc~light curves have smaller amplitudes than \rrab's and, because their nearly sinusoidal light curves are more easily confused with those of other types of variables (e.g. eclipsing contact binaries), they are harder to identify  with fewer epochs.  

For \asas~both the \rrab~and \rrc~completeness maps are remarkably uniform across the sky away from the Galactic plane and show a clear decrease at low Galactic latitudes ($|b|\lesssim25\degr$) at the faintest magnitude bin; plus two areas of lower completeness  than average  around the Magellanic Clouds ($l\sim300\degr$, $b\sim-45$)  and at $l\sim120\degr,b\lesssim 50\degr$ in the second and third magnitude bins. The lower completeness in at low latitude and the Magellanic Clouds is likely due to crowding, since the \asas~telescopes have relatively low angular resolution (pixel size $8"$). The other low completeness area at $l\sim120\degr,b\lesssim 50\degr$ coincides with the North Celestial Pole ($\mathrm{DEC}\lesssim80\degr$), the lower completeness could be due to issues in the \asas~time sampling. By contrast, its uniformity at intermediate to high latitudes for both types of RRLs is a consequence of the large number of epochs per star ($>200$) available in the survey. In these areas, outside the ecliptic plane where \Gaia~is affected by the scanning law, the mean completeness for \Gaia~and \asas~is similar for the \rrab~and slightly lower for the \asas~\rrc~at the faintest two panels and higher for both types in \asas~at the brightest panel (left).

\subsection{Completeness at the faint end: \Gaia~VC+SOS and PS1}\label{s:completeness_ps1}

\begin{figure*}
\begin{center}
\includegraphics[width=2\columnwidth]{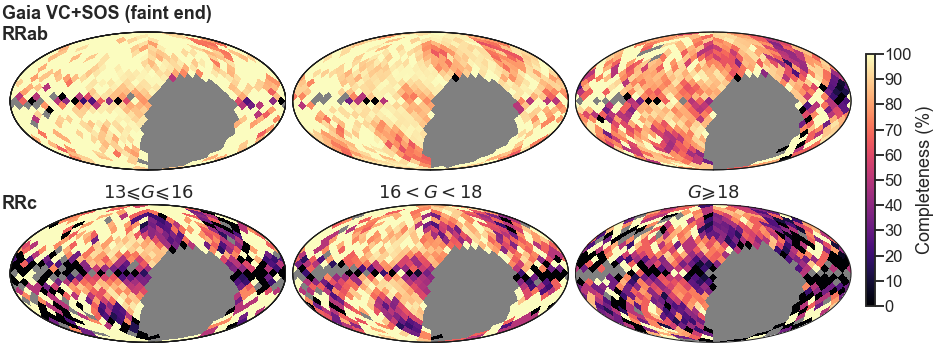}
\caption{Completeness maps for \Gaia~VC+SOS \rrab~(\emph{top}) and \rrc~or \rrd~(\emph{bottom}) stars, estimated from matches against PS1, in three G-band magnitude ranges. All maps are Mollweide projections in Galactic coordinates, with $l=0\degr$ at the centre and longitude increasing to the left. The gray shading denotes pixels with undefined completeness (i.e. zero counts in the comparison survey).}\label{f:gaia_completeness_G_faint_maps}
\end{center}
\end{figure*}

\begin{figure*}
\begin{center}
\includegraphics[width=2\columnwidth]{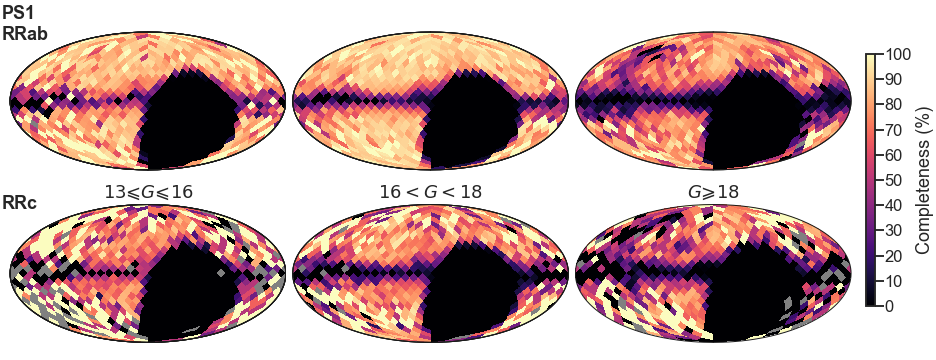}
\caption{Completeness maps for PS1 \rrab~(\emph{top}) and \rrc~or \rrd~(\emph{bottom}) stars, estimated from matches against \Gaia~VC+SOS, in three G-band magnitude ranges. All maps are Mollweide projections in Galactic coordinates, with $l=0\degr$ at the centre and longitude increasing to the left. The gray shading denotes pixels with undefined completeness (i.e. zero counts in the comparison survey).}\label{f:ps1_completeness_G_faint_maps}
\end{center}
\end{figure*}

Following the same procedure as in the previous section, we estimate the completeness of the \Gaia~VC+SOS and PS1 surveys, valid for the magnitude range $13 \leqslant G \leqslant 20.7$. For this we use only the bona fide subset of PS1 RRLs, to ensure the lowest contamination, so our method's assumptions remain valid. The possible loss of completeness due to this cut has no effect in our method.

Figures~\ref{f:gaia_completeness_G_faint_maps} and \ref{f:ps1_completeness_G_faint_maps} show the completeness maps for VC+SOS and PS1 respectively, in each figure separated by RRL subtype. \Gaia~VC+SOS completeness for \rrab~is slightly lower overall at the faint end ($G>18$) and the effect of the scanning law is stronger, but of the order of the large scale variations. For \rrc~stars, the scanning law pattern is visible across all magnitude ranges.  
For the \rrab, the difference of under-sampled patches is $\sim10-20\%$ below the average completeness of $\sim90\%$. For the \rrc, the difference is larger, $\sim30\%$ or higher, as with \asas~reflecting the well-known fact that the lower-amplitude \rrc~stars are more difficult to identify with fewer epochs. 
For PS1 all completeness maps are remarkably spatially homogeneous for both RRL types down to Galactic latitudes $|b|\sim20\degr$, below which there is an significant --and expected-- decline of the completeness towards the Galactic mid-plane. 

\begin{figure}
\begin{center}
\includegraphics[width=\columnwidth]{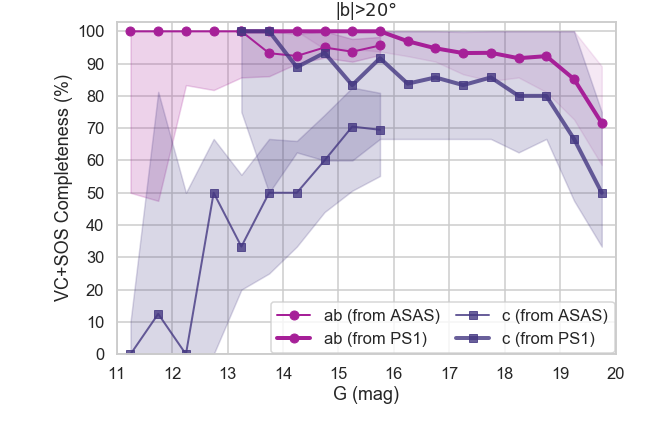}
\includegraphics[width=\columnwidth]{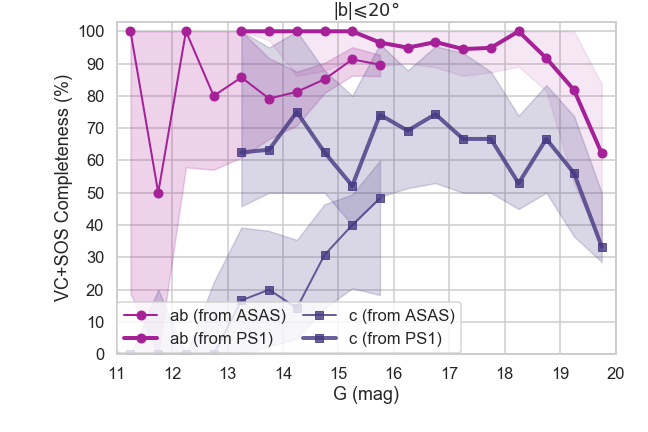}
\caption{Median completeness versus (intensity-averaged) G stars estimated for \Gaia~VC+SOS based on \asas~(\emph{thin}) and on PS1 (\emph{thick}) for \rrab~($\circ$) and \rrc~($\sq$).
The shaded areas represent the completeness' inter-quartile range for the VC+SOS maps. \emph{Top}:  $|b|>20\degr$. \emph{Bottom}: $|b|\leqslant20\degr$.}\label{f:completeness_gaia_vsG_asas_ps1}
\end{center}
\end{figure}

The overlap of the \asas~and PS1 surveys in the magnitude range $13 \lesssim G \lesssim 16$ also allows us to compare the two independent results derived for \Gaia~VC+SOS. The mean completeness across the sky shown in the left maps of Figure~\ref{f:gaia_completeness_G_faint_maps} can be compared to those in the right maps of Figure~\ref{f:gaia_completeness_G_bright_maps}, which confirms the results derived from the two surveys look consistent overall. For a more quantitative comparison, Figure~\ref{f:completeness_gaia_vsG_asas_ps1} shows the median completeness estimates for \Gaia~VC+SOS based on \asas~(thin) and PS1 (thick), again separately for high (top) and low (bottom) latitude fields. To make the comparison straight-forward the results shown correspond to the $3/4$ of the sky in the PS1 footprint (i.e. the area covered by both surveys).  

Over the full magnitude range the two independent estimates show remarkable agreement for the \rrab~stars: in both latitude ranges, the two estimates differ by only a few percent ($<5\%$) and are well within the typical variation observed for different lines of sight (shading). For the \rrc, the estimates from \asas~are systematically lower compared to those from PS1, with differences that can reach $\sim30\%$. This is, again, a result of \rrc~surveys being more prone to be contaminated by other types of variables. Nevertheless, the observed differences are within the typical line-of-sight variations, with the largest discrepancies occurring either at high latitude or at the ends of the two curves where the effect of Poisson noise is expected to be more important given the lower number of stars. 

\subsection{Final Completeness for \Gaia~VC+SOS}\label{s:completeness_full}

\begin{figure*}
\begin{center}
\includegraphics[width=2\columnwidth]{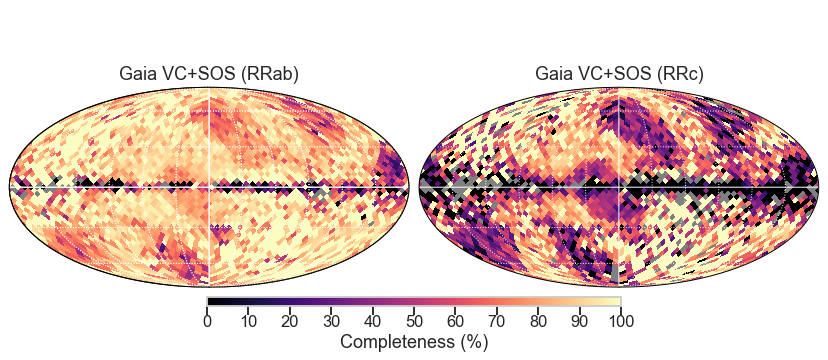}
\caption{Final completeness maps for  Gaia VC+SOS \rrab~(top) and \rrc~(bottom) stars, valid in the full magnitude range of the survey. Maps are Mollweide projections in Galactic coordinates, with $l=0\degr$ at the centre and longitude increasing to the left. The gray shading denotes pixels with \emph{indefinite} completeness (i.e. zero counts in the comparison survey).}\label{f:completeness_gaia_final}
\end{center}
\end{figure*}

So far we have shown that our separate estimates of the VC+SOS completeness based on two mutually independent catalogues, PS1 and \asas, are consistent with one another in the magnitude range of overlap ($13<G<16$) between the two. Therefore, we can provide a final estimate of the completeness over the full magnitude range using the combined PS1 and \asas~catalogues. We compute this taking $A$ as \Gaia~ and $B$ as PS1$\cup$ASAS~in Eq.~(\ref{e:Cg}), making sure objects are not counted twice in the latter.

At the faint end ($G>16$) there remains, however, an area of $\pi$ in the sky ($\mathrm{DEC}<-30\degr$) where the lack of PS1 coverage prevents us from estimating the completeness. To circumvent this and be able to provide an all-sky map, we choose to report the estimated completeness corresponding to the diametrically opposing field in ecliptic coordinates. This is a reasonable approximation because it preserves the symmetry of the completeness pattern with respect to the ecliptic plane, inherited from the \Gaia~scanning law, and because the `hole' in the PS-1 footprint ($\mathrm{DEC}<-30\degr$) corresponds mostly to intermediate to high ecliptic latitudes ($|\beta|>20\degr$) where the Gaia scanning law --and the completeness map-- varies fairly smoothly. This choice also tries to preserve symmetry with respect to the Galactic Plane. So, for a given line of sight $(\lambda_\circ,\beta_\circ)$ we assign the completeness value estimated for the symmetric field at $(\lambda_\circ+180\degr,-\beta_\circ)$. We do this substitution in the full G-band range for consistency.
We caution, however, that our estimate of the completeness from symmetric regions is likely to be over estimated around the Magellanic Clouds, judging from previous studies based on comparisons against the OGLE-IV catalogue (see Sec.~\ref{s:lit_comparison}.

Figure~\ref{f:completeness_gaia_final} shows the final completeness maps for \Gaia~VC+SOS \rrab~(left) and \rrc~(right) for the full G-band magnitude range $11 \leqslant G \leqslant 20.7$. The substitution made in ecliptic coordinates is only hardly noticeable in a couple of slightly lower completeness bits next to some of the lower completeness patches along the ecliptic plane (e.g. $l\sim 0\degr,b\sim-30\degr$ and $l\sim30\degr,b<-60\degr$ for the \rrab). Figure~\ref{f:completeness_gaia_final} shows the final completeness maps for \rrab~(top) and \rrc~(bottom) in three G-band magnitude ranges. 

The completeness maps for \Gaia~VC+SOS, PS1 and \asas, by magnitude range and in full, are publicly available at the GitHub repository \verb+rrl_completeness+\footnote{\href{https://github.com/cmateu?tab=repositories available upon acceptance}{https://github.com/cmateu?tab=repositories}} and as ASCII tables in Appendix~\ref{a:maps}.

\subsection{Completeness as a function of distance and magnitude}

\begin{figure*}
\begin{center}
\includegraphics[width=2\columnwidth]{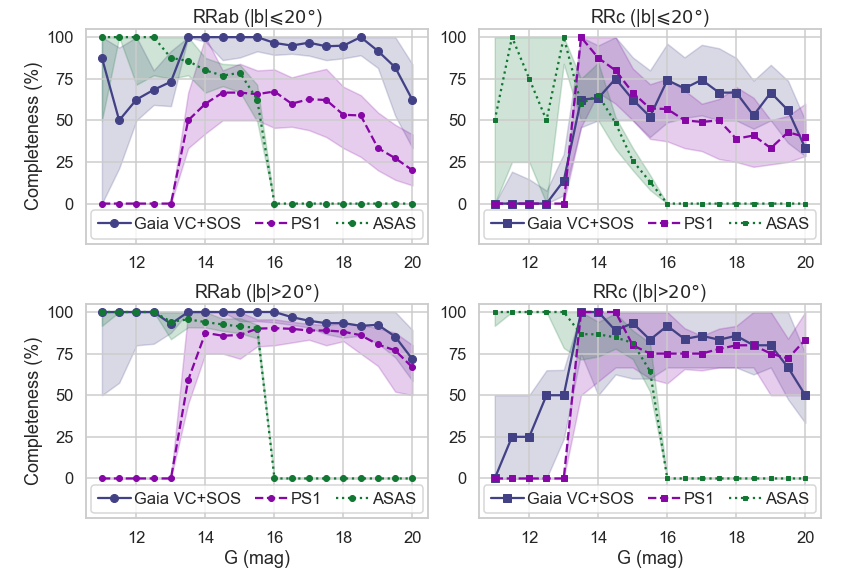} 
\caption{Median completeness for \rrab~(left) and \rrc~(right) stars versus (intensity-averaged) G magnitude for \Gaia~VC+SOS (solid), PS1 (dashed) and \asas~(dotted) RRLs at  $|b|\leqslant20\degr$ (\emph{top}) and $|b|<20\degr$ (\emph{bottom}). The shaded areas represent the completeness' inter-quartile range for each survey, to illustrate the combined effect of the uncertainties due to stochastic noise and line-of-sight variations in completeness.}\label{f:completeness_gaia_ps1_asas_G}
\end{center}
\end{figure*}

Figure~\ref{f:completeness_gaia_ps1_asas_G} and Table~\ref{t:completeness} summarise the median completeness for VC+SOS, PS1 and \asas~as a function of the (intensity-averaged) G-band magnitude, at low ($|b|\leqslant20\degr$) and high ($|b|>20\degr$) Galactic latitudes.

At the bright end ($G\leqslant13$), for \rrab~stars \asas~and \Gaia~ are both near 100\% complete at high latitude; at low latitude \asas~has a higher completeness. For \rrc~stars \asas~has the best performance, with \Gaia's~completeness dropping fast for $G<13$. Note, though, that in this magnitude range, for both surveys, the stochastic noise is notoriously high.  The higher completeness for \asas~is not entirely surprising, at least for the RRLs of type~\typec, since identification of variables in \asas~is based on several hundreds of epochs, many more than \Gaia~in it's current release, and as we have mentioned \rrc~stars are more difficult to identify and confidently tell apart from other variables, particularly with few epochs. 

At the faint end ($G>13$) \Gaia~VC+SOS is more complete than both PS1 and \asas, in all cases. For \rrab~stars the three surveys are highly complete $>85\%$ at high latitude, were all surveys are expected to have the best performance. The differences become more noticeable for the \rrc~stars and at low latitude for both types. Overall, the \Gaia~VC+SOS catalogue has a remarkably high completeness for \rrab, being $>90\%$ up to $G\sim19$ both at high and low latitudes; and lower --as expected-- for \rrc, the difference being more pronounced at low latitudes, where the completeness of the \rrc~drops to $\sim70\%$ compared to $>80\%$ at high latitude. The inter-quartile range is also shown in the figure, to illustrate the combined effect of the stochastic noise and the dispersion of completeness values over the sky for the two types of RRL. This clearly shows the dispersion is more than double for \rrc~stars ($\gtrsim20\%$) than it is for \rrab~stars ($<10\%$) over the entire magnitude (or distance) range, as first observed in the mean maps of Figure~\ref{f:gaia_completeness_G_faint_maps}.    

\begin{figure*}
\begin{center}
\includegraphics[width=\columnwidth]{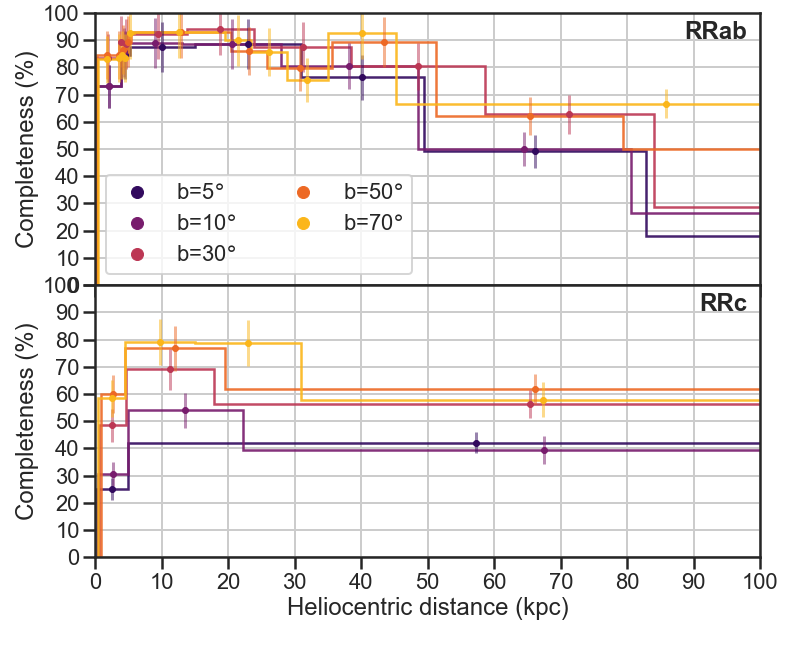}
\includegraphics[width=\columnwidth]{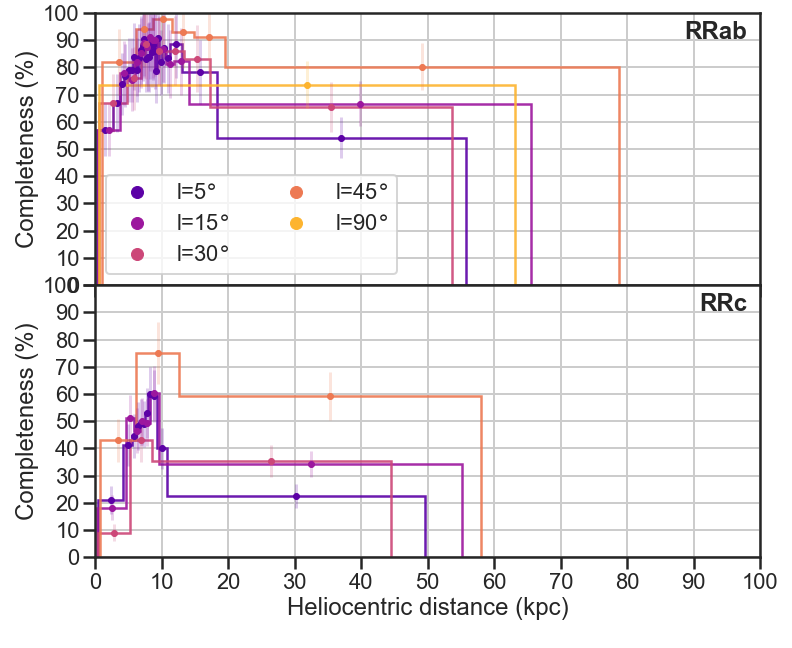}
\caption{Completeness of \Gaia~VC+SOS RRLs as a function of (heliocentric) distance for different lines of sight. \emph{Left:} $l=180\degr$ at different Galactic latitudes, with a fixed angular radius of $30\degr$. \emph{Right:} $b=0\degr$ at different Galactic longitudes, with a fixed angular radius of $10\degr$. The distance bins have a variable width, selected so as to have a minimum number of 100 stars per bin, corresponding to a Poisson noise of $\sim15\%$. }\label{f:completeness_gaia_final_vsD}
\end{center}
\end{figure*}

In cases where there is zero or constant extinction, Fig.~\ref{f:completeness_gaia_ps1_asas_G} 
trivially gives the dependence of completeness upon distance.
In the Galactic plane, however, extinction varies significantly with distance and with the line of sight. In this case it is useful to conduct the same analysis as we have done so far, but as a function of distance rather than G magnitude. We remind the reader that, although 'geometric' distances can be calculated using \Gaia~parallaxes for the nearest objects, throughout this work we use only \emph{photometric} distances (see Sec.~\ref{s:distances}). 
As a use case, Figure~\ref{f:completeness_gaia_final_vsD} illustrates the dependence of the completeness for \Gaia~VC+SOS as a function of distance, for lines of sight toward the Galactic Anti-centre ($l=180\degr$) with increasing Galactic latitude (left panel) and in the Galactic Plane ($b=0\degr$) with increasing Galactic longitude. The bins have variable width in distance, so as to ensure a fixed number of stars per bin and, in turn, a fixed (fractional) Poisson noise. Outside the Galactic plane, for the \rrab~there is very little, if any, dependence with Galactic latitude up to $\sim 40$~kpc; for larger distances there is a small dependence, the completeness drops by $\sim10\%$ for lines-of-sight a $b<10\degr$ compared to ones at higher latitude. By contrast, for the \rrc, the completeness depends more strongly on Galactic latitude at all distances. Even at distances $<20$~kpc the completeness drops from $80\%$ at $b=70\degr$ to $40\%$ at $b\sim5\%$.

The GitHub repository \verb+rrl_completeness+\footnote{\href{https://github.com/cmateu?tab=repositories available upon acceptance}{https://github.com/cmateu?tab=repositories}} provides a convenience function in Python to retrieve completeness as a function of distance for a survey in a given line of sight. Also, for convenience,  completeness as a function of distance, per line-of-sight, for each of the three surveys is provided in Table~\ref{a:3dmaps},  Appendix~\ref{a:maps}.

\begin{table}
\begin{center}
\caption{Median Completeness for VC+SOS, PS1 and \asas~RRLs at $|b|>20\degr$}\label{t:completeness}
\begin{tabular}{lrrrrrr}
\toprule
{G} & \multicolumn{2}{c}{VC+SOS} & \multicolumn{2}{c}{PS1}
& \multicolumn{2}{c}{\asas}\\
{(mag)} &  RRab &  RRc &  RRab &  RRc & RRab &  RRc\\
\midrule
11.0-11.5 &        100.0 &         0.0 &       0.0 &      0.0 &      100.0 &     100.0 \\
11.5-12.0 &        100.0 &        25.0 &       0.0 &      0.0 &      100.0 &     100.0 \\
12.0-12.5 &        100.0 &        25.0 &       0.0 &      0.0 &      100.0 &     100.0 \\
12.5-13.0 &        100.0 &        50.0 &       0.0 &      0.0 &      100.0 &     100.0 \\
13.0-13.5 &         92.9 &        50.0 &       0.0 &      0.0 &       94.2 &     100.0 \\
13.5-14.0 &        100.0 &       100.0 &      59.2 &    100.0 &       95.6 &      86.7 \\
14.0-14.5 &        100.0 &       100.0 &      87.5 &    100.0 &       93.9 &      86.6 \\
14.5-15.0 &        100.0 &        88.9 &      85.7 &    100.0 &       92.5 &      84.7 \\
15.0-15.5 &        100.0 &        93.3 &      86.2 &     80.0 &       91.3 &      81.4 \\
15.5-16.0 &        100.0 &        83.3 &      90.0 &     75.0 &       90.5 &      64.5 \\
16.0-16.5 &        100.0 &        91.7 &      90.3 &     75.0 &        0.0 &       0.0 \\
16.5-17.0 &         96.9 &        83.8 &      89.9 &     75.0 &        0.0 &       0.0 \\
17.0-17.5 &         94.7 &        85.7 &      88.9 &     75.0 &        0.0 &       0.0 \\
17.5-18.0 &         93.3 &        83.3 &      88.9 &     77.8 &        0.0 &       0.0 \\
18.0-18.5 &         93.3 &        85.7 &      88.2 &     80.0 &        0.0 &       0.0 \\
18.5-19.0 &         91.7 &        80.0 &      86.0 &     80.0 &        0.0 &       0.0 \\
19.0-19.5 &         92.3 &        80.0 &      80.5 &     75.0 &        0.0 &       0.0 \\
19.5-20.0 &         85.2 &        66.7 &      76.9 &     72.1 &        0.0 &       0.0 \\
\bottomrule
\end{tabular}

\end{center}
\end{table}

\begin{table}
\begin{center}
\caption{Median Completeness for VC+SOS, PS1 and \asas~RRLs at $|b|\leqslant20\degr$}\label{t:completeness_bleq20}
\begin{tabular}{crrrrrr}
\toprule
{G} & \multicolumn{2}{c}{VC+SOS} & \multicolumn{2}{c}{PS1}
& \multicolumn{2}{c}{\asas}\\
{(mag)} &  RRab &  RRc &  RRab &  RRc & RRab &  RRc\\
\midrule
11.0-11.5 &         87.5 &         0.0 &       0.0 &      0.0 &      100.0 &      50.0 \\
11.5-12.0 &         50.0 &         0.0 &       0.0 &      0.0 &      100.0 &     100.0 \\
12.0-12.5 &         62.5 &         0.0 &       0.0 &      0.0 &      100.0 &      75.0 \\
12.5-13.0 &         68.3 &         0.0 &       0.0 &      0.0 &      100.0 &      50.0 \\
13.0-13.5 &         73.2 &        13.8 &       0.0 &      0.0 &       87.5 &     100.0 \\
13.5-14.0 &        100.0 &        62.5 &      50.0 &    100.0 &       85.7 &      60.0 \\
14.0-14.5 &        100.0 &        63.3 &      60.0 &     87.5 &       80.0 &      65.2 \\
14.5-15.0 &        100.0 &        75.0 &      66.7 &     80.0 &       76.9 &      48.4 \\
15.0-15.5 &        100.0 &        62.5 &      66.7 &     66.7 &       78.6 &      25.7 \\
15.5-16.0 &        100.0 &        52.2 &      65.7 &     57.1 &       62.5 &      13.1 \\
16.0-16.5 &         96.5 &        74.2 &      67.3 &     56.8 &        0.0 &       0.0 \\
16.5-17.0 &         94.9 &        69.2 &      60.0 &     50.0 &        0.0 &       0.0 \\
17.0-17.5 &         96.7 &        74.3 &      62.7 &     49.1 &        0.0 &       0.0 \\
17.5-18.0 &         94.5 &        66.7 &      62.2 &     50.0 &        0.0 &       0.0 \\
18.0-18.5 &         94.9 &        66.7 &      53.3 &     38.9 &        0.0 &       0.0 \\
18.5-19.0 &        100.0 &        52.9 &      52.9 &     40.9 &        0.0 &       0.0 \\
19.0-19.5 &         91.7 &        66.7 &      33.3 &     33.3 &        0.0 &       0.0 \\
19.5-20.0 &         81.8 &        56.1 &      27.3 &     42.9 &        0.0 &       0.0 \\
\bottomrule
\end{tabular}
\end{center}
\end{table}

\subsection{Comparison with the literature}\label{s:lit_comparison}

Completeness rates of the RRLs identified in \Gaia~DR2 were estimated also in other works that validated such candidates. \citet{Molnar2018} compared the RRL candidates in the \Gaia~DR2 classification (VC) with the ones found in a selection of fields of the \textit{Kepler} and \textit{K2} missions \citep{Kepler, K2}. These fields covered about 110~deg$^2$ each and were distributed along the Ecliptic (except for the original \textit{Kepler} field); they spanned Galactic latitudes $|b|<70\degr$ but excluded the region of the Galactic bulge \citep[as already well resolved in][]{OGLE-bulge}. The average completeness rates of the RRL candidates (of all sub-types) in VC were found typically between 70 and 78\%, depending on the data sets, in agreement with the results as a function of sky location illustrated in Fig.~\ref{f:completeness_gaia_final}.

Further comparisons by \citet{Holl2018} of the RRL identifications (all types combined) in \Gaia~DR2 versus OGLE-IV \citep{OGLE-bulge, OGLE-MC} and Catalina \citep{Catalina} estimated average completeness rates between 49 and 63\%. A higher identification rate is expected from OGLE-IV because of its \textit{I}~band (better suited to the reddened objects in the bulge than the \textit{G} band) and the significantly higher sampling with respect to \Gaia. Besides the effect of extinction in the bulge, the distance to the Magellanic Clouds implied lower completeness rates at the faint end, which are consistent with the range covered by the shaded region in grey in Fig.~\ref{f:completeness_gaia_ps1_asas_G}.  The 63\% estimated from Catalina is due to the survey's footprint. Catalina's footprint ($|b|>20\degr$ and $-20\degr<\mathrm{DEC}<70\degr$) covers the ecliptic
plane entirely but has holes at high ecliptic latitude; this biases the mean  towards lower values since areas were \Gaia~has higher completeness (ecliptic poles) are under-represented in the survey. Repeating our analysis using Catalina, instead of \asas, as a reference catalogue we find the median completeness of the healpix map as Fig.~\ref{f:gaia_completeness_G_bright_maps} is 68\%, and a full-sky average of 62\% consistent with the findings from \citet{Holl2018}. As a further check, we also computed the median of our \Gaia~completeness estimated from \asas, but restricting the latter to the Catalina footprint. In this test we find a median completeness of $\sim75\%$. This seems to be in reasonable agreement with the 68\% found from Catalina when considering the difference in magnitude distributions and completeness of the two surveys.

\citet{Rimoldini2019} reported completeness rates of VC as a function of minimum classification score. The average completeness of RRL candidates (of all sub-types, scores, magnitudes, and locations) was estimated at about 65\%, a lower value than expected but likely biased by the large number of OGLE-IV RRL stars (in the bulge and Magellanic Clouds) used in their comparison set.

We can also compare the completeness we have estimated for PS1 with the estimate \citet{Sesar2017b} made based on simulations. \citet{Sesar2017b} find, for the bona fide RRL sample at $|b|>20\degr$, an expected average completeness of 92\% and 79\% for \rrab~and \rrc~stars respectively, up to a distance of 40~kpc ($G\sim18.5$), in excellent agreement with the averages resulting from Table~\ref{t:completeness} for $G<19$, which gives 90.7\% and 82\% for \rrab~and \rrc~respectively. At 80~kpc they estimate an expected completeness of 80\% for \rrab, also in very good agreement with our slightly higher value of $\sim84$\%. At low latitude, however, \citet{Sesar2017b} are not able to provide an estimate for the completeness based on simulations. The empirical approach used here, has allowed us to provide this estimate of PS1's RRL completeness at low galactic latitude for the first time. We'd also like to highlight that the fact the PS1 RRL search was conducted and published in full and not truncated at low latitude  has allowed for our estimate of \Gaia's completeness in its full magnitude range, even close to the Galactic plane, which will be key for upcoming studies of RRLs~in the disc.

\section{Conclusions}\label{s:conclusions}

In this work we have set out to provide a detailed characterisation of the completeness for the three largest RRL~surveys available to date: \Gaia~DR2, PS1 and \asas.

Since \Gaia~DR2 has two partially-overlapping catalogues --SOS and VC-- containing RRLs, we began by validating the combined \Gaia~DR2 VC+SOS catalogue by comparing it against PS1 and other variable star surveys. The main results of our validation process are the following:

\begin{itemize}
    \item Most (81\%) of the RRLs in common between \Gaia~VC+SOS and PS1 are stars classified as bona fide RRL in PS1, a characterisation made by \citep{Sesar2017b} imposing a threshold on the RRL classification scores. The statistics of cross-matches against VC+SOS confirms the PS1 bona fide sample is indeed highly pure and highly complete. Conversely, only a small fraction of the non-bona fide RRL in PS1 ($\sim7\%$) have an RRL counterpart in \Gaia, which confirms only a small fraction of these are likely to be real RRLs.

    \item The period recovery rate of \Gaia~SOS was estimated by cross-matching against PS1 RRLs. When compared to PS1 bona fide RRL, this was found to be 77\% within 3-sigma and about 83\% within 10-sigma. The occurrence of period aliases is found to be small (8--9\%), with PS1 most likely being more affected than \Gaia, as the most common aliases correspond to a 1~d external periodicity.

    \item The contamination of \Gaia~VC+SOS RRLs was estimated at $\sim6$\% on average, based on cross-matches against the \asas~ and CRTS catalogues of periodic variable stars. Separated by RRL~type, contamination is found to be 5\% and 9\% for VC+SOS \rrab~and \rrc\ respectively.
        
\end{itemize}

To estimate the completeness of the RRL surveys we used a probabilistic analysis of stars in common between the different catalogues, following the procedure from \citet{Rybizki2018} and described in Sec.~\ref{s:completeness}. This enabled us to provide completeness estimates for all three catalogues, without assumptions on the completeness of any. The main results of our completeness assessment are summarised as follows:

\begin{itemize}
    \item Completeness of the \Gaia~DR2 VC+SOS RRL catalogue is remarkably high with a median $\gtrsim 95$\% and $>85$\% for \rrab~and \rrc~respectively, even down to the survey limit where it decreases to $\sim85\%$ and $\sim70$\% for \rrab~and \rrc~respectively. Line-of-sight variations in completeness around the ecliptic plane ($\beta<25\degr$) due to the \Gaia~scanning law are $\sim10\%$ for \rrab~and $20-30\%$ for \rrc. These variations affect only \rrab~stars fainter than $G\sim15$ and \rrc~stars in all magnitude ranges.
    Outside the ecliptic plane, \Gaia's completeness is approximately uniform for both RRL types. 
    
    \item \Gaia~DR2 VC+SOS's completeness,  remarkably, shows no significant dependence with Galactic latitude even close to the Galactic Plane for \rrab. There is a mild dependence with Galactic latitude observed for the \rrc, but the main limitation in this estimate is its larger stochastic noise due to the low number of RRLs in the comparison catalogue close to the Galactic Plane.

    \item \asas~RRL's completeness is fairly uniform for both RRL types over the sky and as a function of magnitude outside the Galactic Plane, with an estimated median completeness of 87\% and 58\% for \rrab~and \rrc~ respectively.  At low Galactic latitude $|b|<20\degr$ the completeness rapidly falls off for $G\gtrsim14$. Both \rrab~and \rrc~ maps  show lower completeness than average at the celestial poles at the faint end of the survey $15\leqslant G\leqslant 17$.

    \item PS1 RRL's completeness is remarkably uniform over the sky for both RRL types, and drops sharply to zero when approaching the Galactic Plane at $|b|<20\degr$. At high latitude ($|b|>20\degr$) PS1's median completeness is estimated at 91\% and 82\% down to $G\sim18$ for \rrab~and \rrc respectively; and drops to 76 and 72\% respectively at the faint end ($G\sim20$).

    \item We provide the first estimate of PS1's RRL completeness at low galactic latitude ($|b|\leqslant 20\degr$) estimated at a median $65\%$ and $50-60\%$ for \rrab~and \rrc~stars respectively, up to $G\sim18$ ($g\sim 18.5$).
    
    \item At the bright end ($G\leqslant13$) \asas~has a higher completeness than \Gaia~VC+SOS, except for \rrab~at high latitudes ($|b|>20\degr$) where the two catalogues are near 100\% complete (Fig.~\ref{f:completeness_gaia_ps1_asas_G}). Therefore, \emph{when high completeness at the bright end is desired, supplementing VC+SOS with \asas~RRLs is the optimal strategy}. The combined catalogue will be close to 100\% complete all the way down to $G\sim19$ even close to the Galactic plane.

    \item At the faint end ($G>13$) \Gaia~VC+SOS is more complete than both PS1 and \asas, in all cases. \asas~is also more complete than PS1 down to $G\sim15$, after which completeness rapidly plummets for \asas.
    
    \item The completeness maps shown here for \Gaia~VC+SOS, PS1 and \asas~are provided in Appendix~\ref{a:maps} and are publicly available at the \verb+rrl_completeness+ GitHub repository\footnote{\href{https://github.com/cmateu?tab=repositories}{https://github.com/cmateu?tab=repositories}}, which includes convenience functions in Python to retrieve and compute 2D completeness maps and completeness as a function of distance for a given line of sight.
\end{itemize}{}

This work has been possible thanks to the unprecedented availability of large \emph{independent} surveys of RRL stars overlapping in footprint and depth. The approach followed here can, of course, be applied to any type of tracer survey, and will be particularly valuable for other types of variable star or standard candle tracer. It will also be of enormous use to characterise other large surveys already available, e.g. CRTS and  VVV, and others soon to come, like the variable star surveys to be made by the Vera Rubin Observatory\footnote{formerly the Large Synoptic Survey Telescope (LSST).}.

\section*{Acknowledgements}

CM thanks R. Sanderson for useful discussions. This project was started at the 2018 NYC Gaia Sprint, hosted by the Center for Computational Astrophysics of the Flatiron Institute in New York City. CM thanks the organizers for their encouragement to participate, as well as the Simons Foundation for travel support. JDR thanks the LOC and the SOC for organizing an excellent meeting. Part of this work was performed at the Aspen Center for Physics, which is supported by National Science Foundation grant PHY-1607611. The participation of C.M. at the Aspen Center for Physics was supported by the Simons Foundation. The research leading to these results has received funding from the European Research Council (ERC) under the European Union's Horizon 2020 research and innovation programme (grant agreement N$^\circ$670519: MAMSIE), from the
KU\,Leuven Research Council (grant C16/18/005: PARADISE), as well as from the BELgian federal Science Policy Office (BELSPO) through
PRODEX grants Gaia and PLATO.
This work has made use of data from the European Space Agency (ESA) mission {\it Gaia} (\url{https://www.cosmos.esa.int/gaia}), processed by the {\it Gaia} Data Processing and Analysis Consortium (DPAC,
\url{https://www.cosmos.esa.int/web/gaia/dpac/consortium}). Funding for the DPAC has been provided by national institutions, in particular the institutions participating in the {\it Gaia} Multilateral Agreement. CM acknowledges support from the DGAPA/UNAM PAPIIT program  grant IG100319.

\bibliographystyle{mnras}
\bibliography{cm}



\appendix

\section{Completeness maps for RR Lyrae}\label{a:maps}

\subsection{2D Completeness maps for RR Lyrae}\label{a:2dmaps}

Tables~\ref{t:comp_2d_ab_bright} to \ref{t:comp_2d_c_faint} summarise the 2D completeness maps for RRLs of type~\typeab\ and \typec~presented in Figures~ \ref{f:gaia_completeness_G_bright_maps},\ref{f:asas_completeness_G_bright_maps},  \ref{f:gaia_completeness_G_faint_maps},\ref{f:ps1_completeness_G_faint_maps} and \ref{f:completeness_gaia_final}. The last three columns present the completeness map in each survey's full magnitude range. For PS1 and \asas~the completeness map was computed using \Gaia~DR2 VC+SOS as the reference catalogue. For \Gaia~it was computed using \asas~at the bright end and PS1 at the faint end as described in Sec.~\ref{s:completeness_asas} and \ref{s:completeness_ps1}. The healpix indices provided correspond to maps produced with nested ordering.
The tables' short versions shown are provided to illustrate their form and content. The full tables can be found in the electronic version of the journal.

\begin{table}
\caption{2D completeness map for \asas~and \textit{Gaia}~DR2 VC+SOS \rrab~ in three G-band magnitude ranges at the bright end.}\label{t:comp_2d_ab_bright}
\tabcolsep=0.1cm
\begin{footnotesize}
\begin{tabular}{ccc|cc|cc|cc}
\toprule
Healpix & $l$  & $b$  & ASAS & \Gaia & ASAS & \Gaia & ASAS & \Gaia \\
        & ($\degr$) & ($\degr$) & 
        \multicolumn{2}{|c|}{G $\in$ [11,13]} &  
        \multicolumn{2}{|c|}{G $\in$ [13,15]} &  
        \multicolumn{2}{|c|}{G $\in$ [15,17]} \\
\midrule
0 & 45.00 & +78.28 & 1.00 & 1.00 & 1.00 & 0.69 & 0.46 & 0.91 \\
1 & 135.00 & +78.28 & 1.00 & 0.83 & 0.95 & 0.95 & 0.92 & 0.95 \\
2 & 225.00 & +78.28 & 1.00 & 1.00 & 0.95 & 0.95 & 0.88 & 0.82 \\
3 & 315.00 & +78.28 & 1.00 & 1.00 & 0.96 & 0.79 & 0.79 & 0.89 \\
4 & 22.50 & +66.44 & 1.00 & 1.00 & 0.94 & 0.86 & 0.90 & 0.84 \\
$\cdots$ & $\cdots$ & $\cdots$ & $\cdots$ & $\cdots$ & $\cdots$ & $\cdots$ & $\cdots$ & $\cdots$ \\
\bottomrule
\end{tabular}
\end{footnotesize}
\end{table}

\begin{table}
\caption{2D completeness map for \asas~and \textit{Gaia}~DR2 VC+SOS \rrc~ in three G-band magnitude ranges at the bright end.}\label{t:comp_2d_c_bright}
\tabcolsep=0.1cm
\begin{footnotesize}
\begin{tabular}{ccc|cc|cc|cc}
\toprule
Healpix & $l$  & $b$  & ASAS & \Gaia & ASAS & \Gaia & ASAS & \Gaia \\
        & ($\degr$) & ($\degr$) & 
        \multicolumn{2}{|c|}{G $\in$ [11,13]} &  
        \multicolumn{2}{|c|}{G $\in$ [13,15]} &  
        \multicolumn{2}{|c|}{G $\in$ [15,17]} \\
\midrule
0 & 45.00 & +78.28 & 0.50 & 0.50 & 1.00 & 0.62 & 0.24 & 0.65 \\
1 & 135.00 & +78.28 &  &  & 0.89 & 0.89 & 0.80 & 0.75 \\
2 & 225.00 & +78.28 & 1.00 & 0.25 & 1.00 & 0.82 & 0.51 & 0.75 \\
3 & 315.00 & +78.28 &  & 0.00 & 1.00 & 0.60 & 0.43 & 0.78 \\
4 & 22.50 & +66.44 & 1.00 & 0.17 & 1.00 & 0.29 & 0.71 & 0.54 \\
$\cdots$ & $\cdots$ & $\cdots$ & $\cdots$ & $\cdots$ & $\cdots$ & $\cdots$ & $\cdots$ & $\cdots$ \\
\bottomrule
\end{tabular}
\end{footnotesize}
\end{table}

\begin{table*}
\caption{2D completeness map for \asas, PS1 and \textit{Gaia}~DR2 VC+SOS \rrab~ in three G-band faint magnitude ranges and in the full magnitude range.}\label{t:comp_2d_ab_faint}
\tabcolsep=0.1cm
\begin{footnotesize}
\begin{tabular}{ccc|ccc|ccc|ccc|ccc}
\toprule
Healpix & $l$  & $b$  & ASAS & PS1 & \Gaia & ASAS & PS1 & \Gaia & ASAS & PS1 & \Gaia & ASAS & PS1 & \Gaia \\
        & ($\degr$) & ($\degr$) & 
        \multicolumn{3}{|c|}{G $\in$ [13,16]} &  
        \multicolumn{3}{|c|}{G $\in$ [16,18]} &  
        \multicolumn{3}{|c|}{G $\in$ [18,21]} &
        G $\in$ [10,17] & G $\in$ [13,21] & G $\in$ [10,21]\\
\midrule
0 & 45.00 & +87.08 & 1.00 & 0.50 & 1.00 & 0.50 & 1.00 & 0.67 & 0.00 & 0.50 & 1.00 & 1.00 & 0.60 & 0.89 \\
1 & 135.00 & +87.08 & 1.00 & 0.83 & 1.00 & 0.20 & 1.00 & 0.83 & 0.00 & 1.00 & 0.40 & 1.00 & 0.92 & 0.76 \\
2 & 225.00 & +87.08 & 0.50 & 1.00 & 0.67 & 0.40 & 0.80 & 1.00 & 0.00 & 0.67 & 0.67 & 0.75 & 0.75 & 0.92 \\
3 & 315.00 & +87.08 & 0.67 & 1.00 & 1.00 & 0.62 & 1.00 & 1.00 & 0.00 & 0.67 & 1.00 & 0.86 & 0.93 & 1.00 \\
4 & 22.50 & +84.15 & 0.67 & 1.00 & 1.00 & 0.40 & 1.00 & 1.00 & 0.00 & 0.67 & 1.00 & 0.83 & 0.91 & 1.00 \\
$\cdots$ & $\cdots$ & $\cdots$ & $\cdots$ & $\cdots$ & $\cdots$ & 
$\cdots$ & $\cdots$ & $\cdots$ & $\cdots$ & $\cdots$ & $\cdots$ &
$\cdots$ & $\cdots$ & $\cdots$ \\
\bottomrule
\end{tabular}
\end{footnotesize}
\end{table*}

\begin{table*}
\caption{2D completeness map for \asas, PS1 and \textit{Gaia}~DR2 VC+SOS \rrc~ in three G-band faint magnitude ranges and in the full magnitude range.}\label{t:comp_2d_c_faint}
\tabcolsep=0.1cm
\begin{footnotesize}
\begin{tabular}{ccc|ccc|ccc|ccc|ccc}
\toprule
Healpix & $l$  & $b$  & ASAS & PS1 & \Gaia & ASAS & PS1 & \Gaia & ASAS & PS1 & \Gaia & ASAS & PS1 & \Gaia \\
        & ($\degr$) & ($\degr$) & 
        \multicolumn{3}{|c|}{G $\in$ [13,16]} &  
        \multicolumn{3}{|c|}{G $\in$ [16,18]} &  
        \multicolumn{3}{|c|}{G $\in$ [18,21]} &
        G $\in$ [10,17] & G $\in$ [13,21] & G $\in$ [10,21]\\
\midrule
0 & 45.00 & +87.08 & 1.00 & 1.00 & 0.33 & 1.00 & 1.00 & 1.00 & -- & -- & 0.00 & 1.00 & 1.00 & 0.29 \\
1 & 135.00 & +87.08 & 1.00 & 1.00 & 0.67 & -- & -- & -- & 0.00 & 1.00 & 1.00 & 1.00 & 1.00 & 0.80 \\
2 & 225.00 & +87.08 & 1.00 & 0.75 & 1.00 & -- & -- & -- & 0.00 & 0.00 & 0.00 & 1.00 & 0.50 & 0.83 \\
3 & 315.00 & +87.08 & 0.50 & 0.00 & 0.50 & 0.00 & 1.00 & 1.00 & -- & -- & -- & 0.50 & 0.67 & 0.83 \\
4 & 22.50 & +84.15 & 1.00 & 1.00 & 0.50 & 0.00 & 0.67 & 0.67 & 0.00 & 1.00 & 1.00 & 0.50 & 0.80 & 0.67 \\
$\cdots$ & $\cdots$ & $\cdots$ & $\cdots$ & $\cdots$ & $\cdots$ & 
$\cdots$ & $\cdots$ & $\cdots$ & $\cdots$ & $\cdots$ & $\cdots$ &
$\cdots$ & $\cdots$ & $\cdots$ \\
\bottomrule
\end{tabular}
\end{footnotesize}
\end{table*}

\subsection{3D Completeness maps for RR Lyrae}\label{a:3dmaps}

Tables \ref{t:comp_3d_gaia_ab} to \ref{t:comp_3d_asas_c} summarise the completeness maps versus line of sight and distance for RRLs of type~\typeab\ and \typec, in a healpix level 2 grid computed with a line of sight radius of $16\degr$ (approximately corresponding to twice the healpixel's area) and variable-width radial bins to ensure 30 stars per bin. The completeness map for \Gaia~DR2 VC+SOS was computed using the union of \asas~and PS1 and substituting the fields at $\mathrm{DEC}<-30\degr$ with their ecliptic diametrical opposite fields as discussed in Sec.~\ref{s:completeness_full}. For PS1 and \asas~the completeness map was computed using \Gaia~DR2 VC+SOS as the reference catalogue. The healpix indices provided correspond to maps produced with nested ordering.
The tables' short versions shown are provided to illustrate their form and content. The full tables can be found in the electronic version of the journal.

\begin{table}
\caption{3D completeness map for \textit{Gaia}~DR2 VC+SOS \rrab.}\label{t:comp_3d_gaia_ab}
\tabcolsep=0.1cm
\begin{footnotesize}
\begin{tabular}{ccccccc}
\toprule
Healpix & $l$  & $b$  & $D_o$  & $D_f$  &  Completeness & Error  \\
        & ($\degr$) & ($\degr$) & (kpc) & (kpc) &       &        \\
\midrule
0 & 45.00 & +78.28 & 1.3 & 3.9 & 1.00 & 0.26 \\
0 & 45.00 & +78.28 & 3.9 & 5.5 & 0.97 & 0.25 \\
0 & 45.00 & +78.28 & 5.5 & 6.8 & 0.93 & 0.25 \\
0 & 45.00 & +78.28 & 6.8 & 7.9 & 0.97 & 0.25 \\
0 & 45.00 & +78.28 & 7.9 & 9.2 & 0.93 & 0.25 \\
$\cdots$ & $\cdots$ & $\cdots$ & $\cdots$ & $\cdots$ & $\cdots$ & $\cdots$ \\
\bottomrule
\end{tabular}
\end{footnotesize}
\end{table}

\begin{table}
\caption{3D completeness map for \textit{Gaia}~DR2 VC+SOS \rrc.}\label{t:comp_3d_gaia_c}
\tabcolsep=0.1cm
\begin{footnotesize}
\begin{tabular}{ccccccc}
\toprule
Healpix & $l$  & $b$  & $D_o$  & $D_f$  &  Completeness & Error  \\
        & ($\degr$) & ($\degr$) & (kpc) & (kpc) &       &        \\
\midrule
0 & 45.00 & +78.28 & 1.2 & 4.6 & 0.53 & 0.17 \\
0 & 45.00 & +78.28 & 4.6 & 6.8 & 0.57 & 0.17 \\
0 & 45.00 & +78.28 & 6.8 & 9.4 & 0.80 & 0.22 \\
0 & 45.00 & +78.28 & 9.4 & 12.0 & 0.80 & 0.22 \\
0 & 45.00 & +78.28 & 12.0 & 14.3 & 0.70 & 0.20 \\
$\cdots$ & $\cdots$ & $\cdots$ & $\cdots$ & $\cdots$ & $\cdots$ & $\cdots$ \\
\bottomrule
\end{tabular}
\end{footnotesize}
\end{table}

\begin{table}
\caption{3D completeness map for PS1 \rrab.}\label{t:comp_3d_ps1_ab}
\tabcolsep=0.1cm
\begin{footnotesize}
\begin{tabular}{ccccccc}
\toprule
Healpix & $l$  & $b$  & $D_o$  & $D_f$  &  Completeness & Error  \\
        & ($\degr$) & ($\degr$) & (kpc) & (kpc) &       &        \\
0 & 45.00 & +78.28 & 0.7 & 3.6 & 0.23 & 0.10 \\
0 & 45.00 & +78.28 & 3.6 & 5.3 & 0.90 & 0.24 \\
0 & 45.00 & +78.28 & 5.3 & 6.6 & 0.77 & 0.21 \\
0 & 45.00 & +78.28 & 6.6 & 7.7 & 0.80 & 0.22 \\
0 & 45.00 & +78.28 & 7.7 & 9.0 & 0.87 & 0.23 \\
$\cdots$ & $\cdots$ & $\cdots$ & $\cdots$ & $\cdots$ & $\cdots$ & $\cdots$ \\
\bottomrule
\end{tabular}
\end{footnotesize}
\end{table}

\begin{table}
\caption{3D completeness map for PS1 \rrc.}\label{t:comp_3d_ps1_c}
\tabcolsep=0.1cm
\begin{footnotesize}
\begin{tabular}{ccccccc}
\toprule
Healpix & $l$  & $b$  & $D_o$  & $D_f$  &  Completeness & Error  \\
        & ($\degr$) & ($\degr$) & (kpc) & (kpc) &       &        \\
0 & 45.00 & +78.28 & 1.2 & 6.2 & 0.53 & 0.17 \\
0 & 45.00 & +78.28 & 6.2 & 9.3 & 0.57 & 0.17 \\
0 & 45.00 & +78.28 & 9.3 & 10.1 & 0.40 & 0.14 \\
0 & 45.00 & +78.28 & 10.1 & 13.6 & 0.70 & 0.20 \\
0 & 45.00 & +78.28 & 13.6 & 16.0 & 0.77 & 0.21 \\
$\cdots$ & $\cdots$ & $\cdots$ & $\cdots$ & $\cdots$ & $\cdots$ & $\cdots$ \\
\bottomrule
\end{tabular}
\end{footnotesize}
\end{table}

\begin{table}
\caption{3D completeness map for \asas\ \rrab.}\label{t:comp_3d_asas_ab}
\tabcolsep=0.1cm
\begin{footnotesize}
\begin{tabular}{ccccccc}
\toprule
Healpix & $l$  & $b$  & $D_o$  & $D_f$  &  Completeness & Error  \\
        & ($\degr$) & ($\degr$) & (kpc) & (kpc) &       &        \\
0 & 45.00 & +78.28 & 0.7 & 3.6 & 0.87 & 0.23 \\
0 & 45.00 & +78.28 & 3.6 & 5.3 & 1.00 & 0.26 \\
0 & 45.00 & +78.28 & 5.3 & 6.6 & 1.00 & 0.26 \\
0 & 45.00 & +78.28 & 6.6 & 7.7 & 1.00 & 0.26 \\
0 & 45.00 & +78.28 & 7.7 & 9.0 & 0.87 & 0.23 \\
$\cdots$ & $\cdots$ & $\cdots$ & $\cdots$ & $\cdots$ & $\cdots$ & $\cdots$ \\
\bottomrule
\end{tabular}
\end{footnotesize}
\end{table}

\begin{table}
\caption{3D completeness map for \asas\ \rrc.}\label{t:comp_3d_asas_c}
\tabcolsep=0.1cm
\begin{footnotesize}
\begin{tabular}{ccccccc}
\toprule
Healpix & $l$  & $b$  & $D_o$  & $D_f$  &  Completeness & Error  \\
        & ($\degr$) & ($\degr$) & (kpc) & (kpc) &       &        \\
0 & 45.00 & +78.28 & 1.2 & 6.2 & 0.97 & 0.25 \\
0 & 45.00 & +78.28 & 6.2 & 9.3 & 0.77 & 0.21 \\
0 & 45.00 & +78.28 & 9.3 & 10.1 & 0.23 & 0.10 \\
0 & 45.00 & +78.28 & 10.1 & 13.6 & 0.77 & 0.21 \\
0 & 45.00 & +78.28 & 13.6 & 16.0 & 0.43 & 0.14 \\
$\cdots$ & $\cdots$ & $\cdots$ & $\cdots$ & $\cdots$ & $\cdots$ & $\cdots$ \\
\bottomrule
\end{tabular}
\end{footnotesize}
\end{table}

\section{Period recovery criteria}\label{a:perRecovCrit}
To estimate period recovery for an ensemble of objects, it is desirable to use a statistic that has a fixed variance so that all results can be compared on equal footing and counted using a single threshold.
Let us look at two often used statistics: the difference $\Delta f$ and ratio $R$ between two independently measured frequencies of the same object:
\begin{equation}
\begin{split}
\Delta f &= f_1 - f_2 \quad \quad \quad
\sigma^2_{\Delta f} = \sigma^2_{f_1} + \sigma^2_{f_2} \label{eq:freqDiff}\\
R &= f_1/f_2 \quad \quad \quad \ \ \ \
\sigma^2_{R} = R^2 \left[ 
\left( \frac{\sigma_{f_1}}{f_1} \right)^2 +
\left( \frac{\sigma_{f_2}}{f_2} \right)^2 
\right]
\end{split}
\end{equation}
with $\sigma^2_{f_1}$ and $\sigma^2_{f_2}$ being the respective uncertainties on the frequencies $f_1$ and $f_2$. The provided variances $\sigma^2_{\Delta f}$ and $\sigma^2_{R}$ explicitly assume that both frequencies are uncorrelated, i.e., estimated from independently measured time series.
From Eq.~(\ref{eq:freqDiff}) it is clear that a standard-deviation normalised frequency difference $\overline{\Delta f}$ will have fixed variance of 1:
\begin{align}\label{eqNorFreqDiff}
\overline{\Delta f} 
= \frac{f_1 - f_2}{\sigma_{\Delta f}} 
= \frac{f_1 - f_2}{\sqrt{\sigma^2_{f_1} + \sigma^2_{f_2}}}& 
\quad \quad 
\sigma^2_{\overline{\Delta f}} = \frac{\sigma^2_{f_1} + \sigma^2_{f_2}}{\sigma^2_{f_1} + \sigma^2_{f_2}} = 1\\
\mathrm{recovery\ if:} \ \ | \overline{\Delta f} | < X&
\end{align}
where recovery can then be defined as being within $X$-sigma.
It is however not possible to find such fixed-variance expression for the ratio $R$, and its use is therefore discouraged when used in combination with a single threshold criterion (despite its wide use in literature), unless only a very narrow relative frequency range is examined and when frequency errors are all similar.

Depending on the time-sampling of the examined time series, strong aliases might occur, causing the identified frequency to be offset by a certain value and (generally) a phase shift in the folded light curve. If the time-sampling is known, the exact offset can be computed from the spectral window and, importantly, does not enlarge the uncertainty on the frequency. When the time-sampling is not available, but (very) similar for all objects, it can also be estimated from the ensemble. Using the latter method, we confirm that the prominent 1~day (and 0.5~day harmonic) PanSTARRS-1 alias is at the expected sidereal day, i.e., expressed in frequency: 366.25/365.25 = 1.00273 (solar) day$^{-1}$. 

\subsection{Normalised frequency difference from the data}
\begin{figure}
\includegraphics[width=\columnwidth]{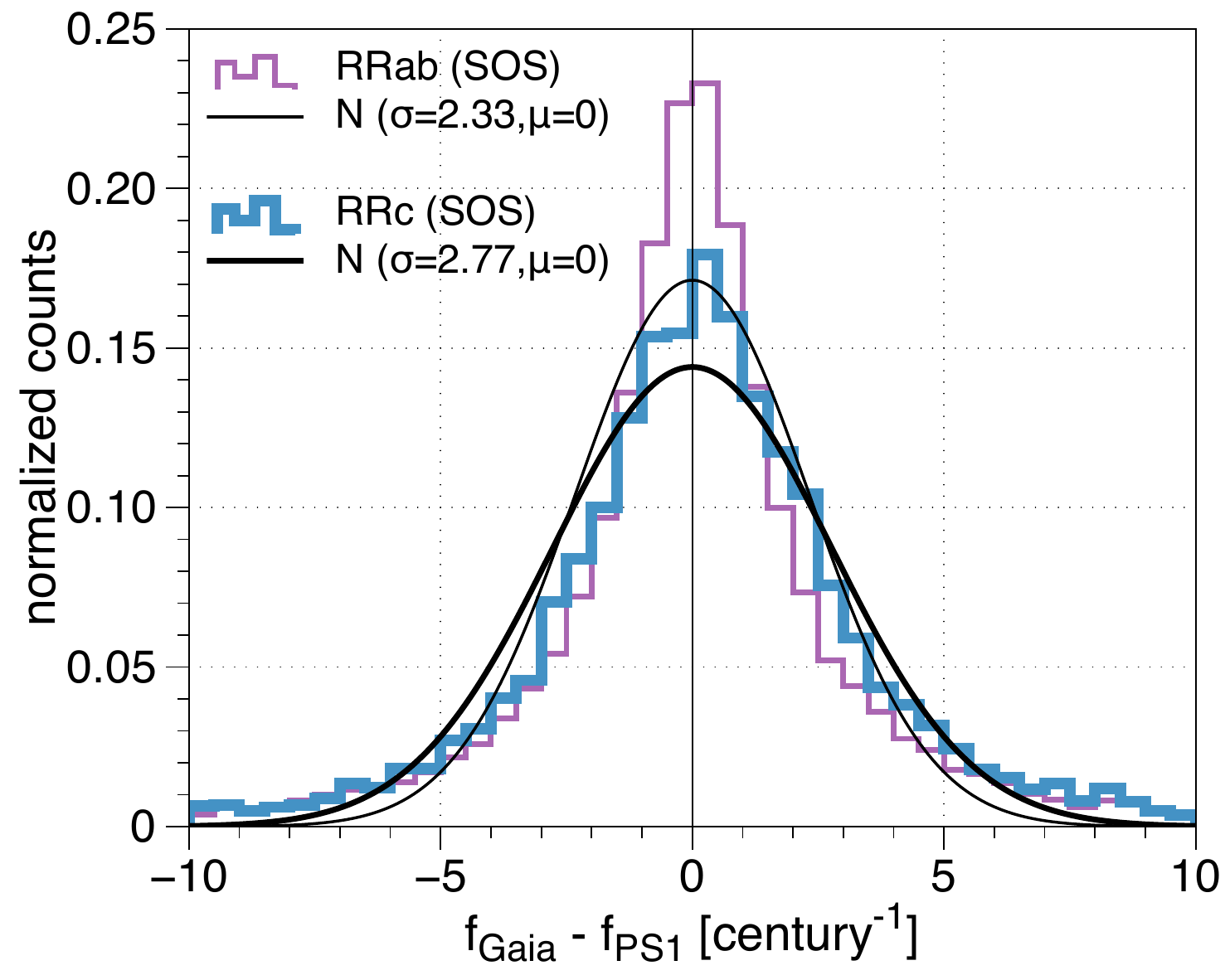}
\includegraphics[width=\columnwidth]{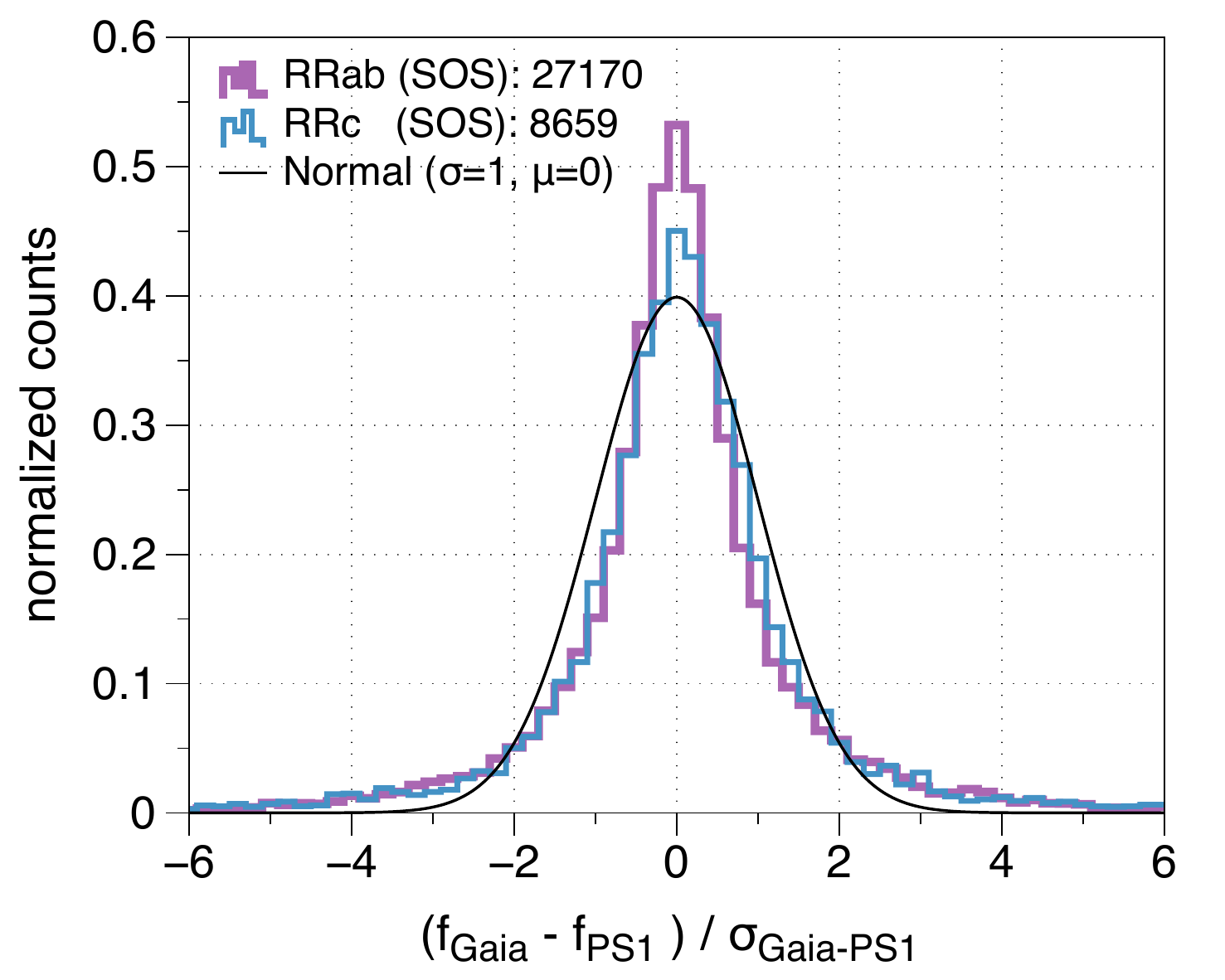}
\caption{
Top panel: frequency difference between \Gaia ~SOS and PS1 around $\Delta f = 0$, i.e., the data directly around the black lines in the bottom two panels of Fig.~\ref{f:period_rec}, which likely is the most alias-free subset of the data. The black lines  show normal distributions $\mathrm{N}(\sigma,\mu)$ derived to represent the data of each type. 
Bottom panel: same after normalisation following the procedure of option 1 (see text). A threshold cut of $| \overline{\Delta f} | < X$ is equal to selecting all sources between [-X,X] in the bottom panel. 
}\label{f:diffFhist}
\end{figure}

Let us start with the distribution of $f_\mathrm{Gaia} - f_\mathrm{PS1}$ as shown in Fig.~\ref{f:diffFhist}. Though the distributions are relatively close to Normal, the heavier tail and more peaked centre suggests that for each type we are dealing with hetroscedastic (non-identical) frequency errors. This is not unexpected as the frequency uncertainty depends on various factors: $\sigma_{f} \propto \sigma_\mathrm{noise} / (A \, T \,  \sqrt{N})$ with $A$ the signal amplitude, $T$ the time series duration and $N$ the number of observations \citep{1980Ap&SS..69..485K,Cuypers1987}, and  the \Gaia ~SOS peak to peak model amplitude ($A$) can vary already with a factor of 2-3. Note that this theory is also the reason we prefer to work with frequency errors: if all factors stay the same the frequency error is (to first order) independent of the actual frequency (or period) of the signal, while the period error will scale as 
$\sigma_p \simeq \sigma_f  f^{-2}$.

Because we intend to use Eq.~(\ref{eqNorFreqDiff}) for our recovery criterion, we ideally require individual frequency uncertainty estimates for our sources.
For the \Gaia ~DR2 RR Lyrae stars \citep{Clementini2018} the uncertainty of each individual period was estimated from a 100-fold Monte-Carlo sampling of the time series \citep[see section 2.1,][]{Clementini2016b}. The PS1 periods estimates in \citet{Sesar2017b}, $p_\text{PS1}$, unfortunately do not have individual error estimates. They do provide a cumulative distribution of $|p_\text{PS1} - p_\mathrm{SDSS}|$ 
where $p_\mathrm{SDSS}$ is the period derived in \citet{Sesar2010} from SDSS Stripe 82. In \citet{Sesar2010} it is mentioned that their period errors are probably around a few seconds, though in \citet{Sesar2017b} it is assumed to be the `true' period, i.e., (much) better than the PS1 derived period.

\begin{figure}
\includegraphics[width=\columnwidth]{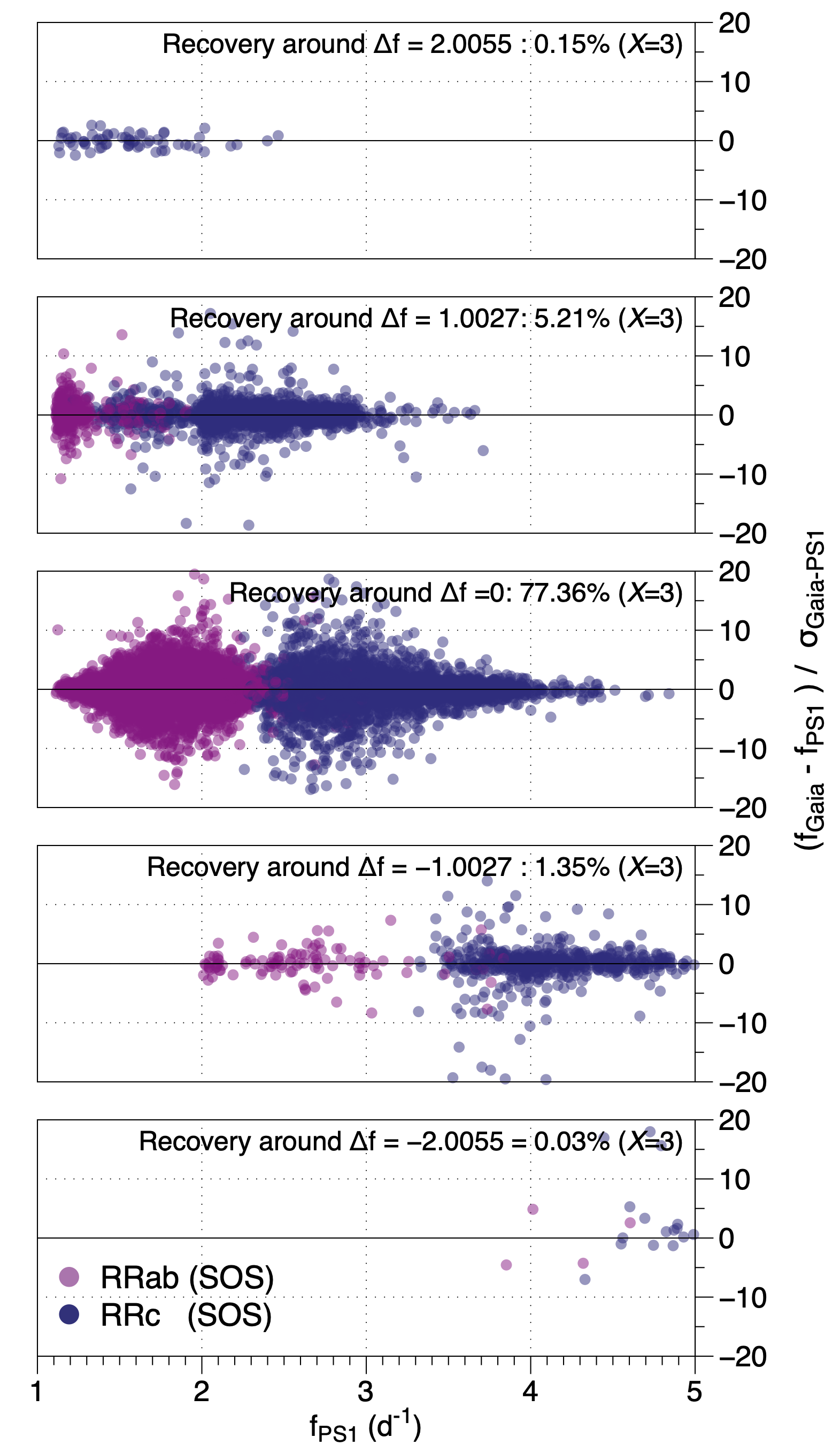}
\caption{Normalized frequency difference between \Gaia ~SOS and PS1 around zero and the main aliases, this corresponds to the option 1 (see text): simply dividing the frequency difference distribution by the best fit standard deviation per type.}\label{f:normFreqDiffDistr}
\end{figure}

The absence of individual PS1 period errors gives us two options: (1)~Simply normalise the distributions of the top panel of Fig.~\ref{f:diffFhist} using its distribution width. For this we use a sigma derived from a fit to the cumulative distribution function\footnote{The statistical standard deviations of the data in this range were 2.83 and 3.08 cycles century$^{-1}$, respectively. But the associated normal distributions are (much) too narrow. Hence we opted to fit the regular gridded CDFs in the indicated range to give more weight to any discrepancy of the area under the curve.} between [-10,10] cycles century$^{-1}$. This means that we simply divide the frequency difference of all the stars in the RRab and RRc samples by  $\sigma_{\Delta f} = 2.33$ and $2.77$ cycles century$^{-1}$, respectively, resulting in the bottom panel of Fig.~\ref{f:diffFhist}. Note that in the figures  $\sigma_{\Delta f}$ is denoted as $\sigma_\mathrm{Gaia-PS1}$ for clarity. The recovery regarding to this criterion is given in Table~\ref{t:period_rec} and shown for the main aliases in Fig.~\ref{f:normFreqDiffDistr}.
For ease of use we here provide the practical formula to compute the unit-less $\overline{\Delta f}$ (using frequency units day$^{-1}$), with $f_\mathrm{alias}$ the estimated mean value of $\Delta f$ of the `alias' that is analysed: 
\begin{align}\label{eq:aliasCorrectedRecovery}
\overline{\Delta f} =&  \frac{f_\mathrm{Gaia} - f_\mathrm{PS1} - f_\mathrm{alias}}{\sigma_{\Delta f} }\\
&\mathrm{with} \ f_\mathrm{alias} = \text{mean $\Delta f$  offset of the `alias'} \nonumber\\
&\mathrm{with} \ \sigma_{\Delta f}=
\begin{cases}
    6.38\times  10^{-5}~\mathrm{day}^{-1}, \text{if RRab} \nonumber \\
    7.58\times  10^{-5}~\mathrm{day}^{-1}, \text{if RRc}
\end{cases} \\
& \text{and recovery if:} \ \ | \overline{\Delta f} | < X& \nonumber
\end{align}
The  $\Delta p$ corresponding to a certain value of threshold $X$ times $|\overline{\Delta f}|$ for period $p$ (in days) can be computed as (with $\sigma_{\Delta f}$ defined above):
\begin{align} \label{f:deltaPeriodEquivalent}
\left| \Delta p \right| \simeq X \ p^{2} \sigma_{\Delta f}
\end{align}

Option (2) is to try to estimate the error distribution for the PS1 periods (as well as \Gaia~ period errors as they might not all be realistic). We have done so by following the example of  \citet{Sesar2017b}: we cross-matched our SOS sample against the \citet{Sesar2010} sources.
For these sources we estimated the width of the distribution in frequency difference of  PS1 (and \Gaia~) against SDSS, like in Fig.~\ref{f:diffFhist}, and used this to derive independent uncertainty estimates. We also tested the estimated PS1 error distribution in combination with the available \Gaia~ SOS per-source uncertainties. In short: any form of re-estimated error distributions did not provide a significantly different frequency recovery with respect to what was found with option 1. Hence we opted to suppress any further details about this alternative procedure and conclude that the simple approach of option 1 is sufficient for this purpose.


\bsp	
\label{lastpage}
\end{document}